\documentclass[9pt, twocolumn]{article}

\usepackage{fullpage}
\usepackage{gensymb}
\usepackage{textcomp}
\usepackage{graphicx}
\usepackage{natbib}

\usepackage{soul}
\usepackage{amsmath,amsfonts,amssymb,bm}
\usepackage[scaled=.97]{helvet} 
\usepackage{mathptmx}
\usepackage[colorlinks=true,allcolors=blue]{hyperref}

\usepackage{url}
\usepackage{doi}

\usepackage{hypernat}

\usepackage[tiny]{titlesec}

\newcommand{\dg}{\textdegree}
\newcommand{\dgs}{\textdegree }
\newcommand{\dgN}{\textdegree N}
\newcommand{\dgNs}{\textdegree N }
\newcommand{\dgS}{\textdegree S}

\newcommand{\dgW}{\textdegree W}
\newcommand{\dgWs}{\textdegree W }
\newcommand{\dgE}{\textdegree E}
\newcommand{\dgEs}{\textdegree E }
\newcommand{\mones}{$^{-1}$ }

\newcommand{\mone}{$^{-1}$}
\newcommand{\mtwo}{$^{-2}$}

\newcommand{\rFigSchematic}[1]{Fig.~\ref{fig:schematic}#1}
\newcommand{\rFigOverview}[1]{Fig.~\ref{fig:overview}#1}
\newcommand{\rFigSHLClim}[1]{Fig.~\ref{fig:shlClim}#1}
\newcommand{\rFigGhtMaps}[1]{Fig.~\ref{fig:ghtMaps}#1}
\newcommand{\rFigGhtCI}[1]{Fig.~\ref{fig:ghtCI}#1}

\newcommand{\rFigHLLocation}[1]{Fig.~\ref{fig:hlLocation}#1}
\newcommand{\rFigVertSections}[1]{Fig.~\ref{fig:vertSections}#1}
\newcommand{\rFigLayerIntDiv}[1]{Fig.~\ref{fig:layerIntDiv}#1}
\newcommand{\rFigLayerDivReg}[1]{Fig.~\ref{fig:layerDivReg}#1}
\newcommand{\rFigLayerDivWRF}[1]{Fig.~\ref{fig:layerDivWrf}#1}
\newcommand{\rFigSfnWRF}[1]{Fig.~\ref{fig:sfnWrf}#1}

\newcommand{\fullWidth}{39pc}
\newcommand{\halfWidth}{19pc}

\title{Weakening and Shifting of the Saharan Shallow Meridional Circulation During Wet Years of the West African Monsoon}

\author{Ravi Shekhar (ravi.shekhar@yale.edu) \\ William R. Boos\\Yale University, New Haven, Connecticut}
\date{January 2017}

\begin{document}
\maketitle
\begin{abstract}
The correlation between increased Sahel rainfall and reduced Saharan surface pressure is well established in observations and global climate models, and has been used to imply that increased Sahel rainfall is caused by a stronger shallow meridional circulation (SMC) over the Sahara. This study uses two atmospheric reanalyses to examine interannual variability of Sahel rainfall and the Saharan SMC, which consists of northward near-surface flow across the Sahel into the Sahara and southward flow near 700 hPa out of the Sahara. During wet Sahel years, the Saharan SMC shifts poleward, producing a drop in low-level geopotential and surface pressure over the Sahara. Statistically removing the effect of the poleward shift from the low-level geopotential eliminates significant correlations between this geopotential and Sahel precipitation. As the Saharan SMC shifts poleward, its mid-tropospheric divergent outflow decreases, indicating a weakening of its overturning mass flux.  The poleward shift and weakening of the Saharan SMC during wet Sahel years is reproduced in an idealized model of West Africa; a wide range of imposed sea surface temperature and land surface albedo perturbations in this model produce a much larger range of SMC variations that nevertheless have similar quantitative associations with Sahel rainfall as in the reanalyses. These results disprove the idea that enhanced Sahel rainfall is caused by strengthening of the Saharan SMC.  Instead, these results are consistent with the hypothesis that the a stronger SMC inhibits Sahel rainfall, perhaps by advecting mid-tropospheric warm and dry air into the precipitation maximum.
\end{abstract}
\section{Introduction}
\label{sec:introduction}
Over the twentieth century, large interannual and interdecadal variations in precipitation were observed in the African Sahel, producing occasional floods and sustained droughts. A variety of studies examined the cause of these variations \citep[e.g.][]{Charney1975, Folland1986, Eltahir1996, Nicholson2001}, but a robust explanation was not established until \citet{Giannini2003} showed that much of the observed variability could be reproduced if observed global SSTs were used to drive a global climate model, implicating SST as the primary cause of historical Sahel precipitation changes.

While it is now generally agreed that SST drives interdecadal variations in Sahel precipitation \citep[e.g.][]{Nicholson2013}, the Sahara desert is also known to be associated with Sahel variability on a range of time scales. \citet{Haarsma2005} found a correlation on interannual time scales between increased Sahel rainfall and decreased mean sea level pressure over the Sahara. They argued that variations in the mean sea level pressure gradient between the Sahara and its surroundings cause variations in low-level  convergence of mass and moisture, and thus in rainfall, over the Sahel. They furthermore argued that the mean sea level pressure gradient is set by the land-ocean temperature contrast, which can then be viewed as a fundamental driver of Sahel rainfall. In contrast, \citet{Biasutti2009} found that land-ocean temperature contrast is poorly correlated with Sahel precipitation at interannual time scales in CMIP3 (Coupled Model Intercomparison Project Phase 3) models.  They argued that variation in Sahel rainfall is caused by fluctuations in the Saharan low, an area of near surface low pressure stretching across the Sahara desert. To quantify this, they calculated the index $\Delta$Z925, defined as the difference in geopotential height at 925~hPa between the Sahara desert and the global tropics. When this index was anomalously low, Sahel precipitation was high. Furthermore, a lead-lag correlation showed this $\Delta$Z925 index was maximally correlated with Sahel precipitation when the index led by one month, suggesting that Saharan Low anomalies cause Sahel rainfall variability. 

The studies just discussed treat the low-level circulation over the Sahara as an entity that can be described in terms of the distribution of mean sea level pressure or 925~hPa geopotential height.  However, this circulation consists of a geopotential height minimum at 925~hPa over the Sahara and a geopotential height maximum in the lower mid-troposphere (near 700~hPa) over the Sahara, with cyclonic and anticyclonic winds rotating around the near-surface geopotential minimum and the 700~hPa geopotential maximum, respectively.  \rFigSchematic{} provides a schematic of this well-known structure of the Saharan circulation \citep[e.g.][]{Thorncroft2011}.  In addition to the balanced cyclonic and anticyclonic flow, mass converges into the near-surface low, ascends, and diverges out of the lower mid-tropospheric high in an ageostrophic overturning circulation.  This shallow overturning, which we refer to as the Saharan shallow meridional circulation (SMC), extends across the entire Sahara desert:  Fig.~\ref{fig:smcSectionLon} clearly shows near-surface poleward flow across the Sahel extending from the west coast of Africa to the East African highlands at nearly 40\dgEs, with return flow at 700 hPa extending over the same region (the data and methods used to create this figure are described in the next section).  Shallow meridional circulations have been documented over Africa \citep{Trenberth2000}, in the Eastern Pacific ocean \citep[e.g.][]{Zhang2004,Nolan2007,Nolan2010}, and in the Australian monsoon \citep{Nie2010}. In continental monsoons such as those in Africa, the ascent branch of the SMC is located poleward of the ascent branch of the deep, precipitating intertropical convergence zone (ITCZ), whereas the ascent in the East Pacific SMC is spatially coincident with the ITCZ \citep{Zhang2008}.

The idea that a stronger Saharan SMC causes increased Sahel precipitation in the summer mean was stated explicitly by \citet{Martin2014a}, who examined differences in the West African monsoon between warm and cold phases of the Atlantic Multidecadal Oscillation (AMO).  They found Sahel precipitation was enhanced during decades when the North Atlantic was warm, and they stated that the Saharan SMC was stronger during the spring (April--June) immediately preceding the summer rainy season (July--Sept.) of those decades.  However, enhanced springtime ascent in their analysis was centered over the Sahel and extended to the upper troposphere, with southward flow out of that ascending region located at 600 hPa and higher altitudes; in contrast, ascent in the Saharan SMC is typically confined below 600 hPa with the southward outflow centered at 700 hPa.  Furthermore, \citet{Hurley2013} found that 700 hPa equatorward outflow from continental SMCs weakened during anomalously rainy years of the West African, southern African, Australian, and South Asian monsoons.  Thus, wet Sahel years seem to be characterized by something other than a simple intensification of the Saharan SMC.  Interpretation of the structures seen in the analysis of \citet{Martin2014a} are complicated by the coarse resolution of the NCEP-NCAR reanalysis \citep{Kalnay1996} that they employed, which has only three to four vertical levels that span the entire SMC overturning.  One of our present goals is to use more recent atmospheric reanalyses to describe the detailed changes in structure and intensity of the Saharan SMC that accompany interannual changes in Sahel precipitation --- interannual variability in the last 30 years is expected to be better constrained by observations than interdecadal variability over the last 100 years.

\begin{figure}
\centering
\includegraphics[width=\halfWidth]{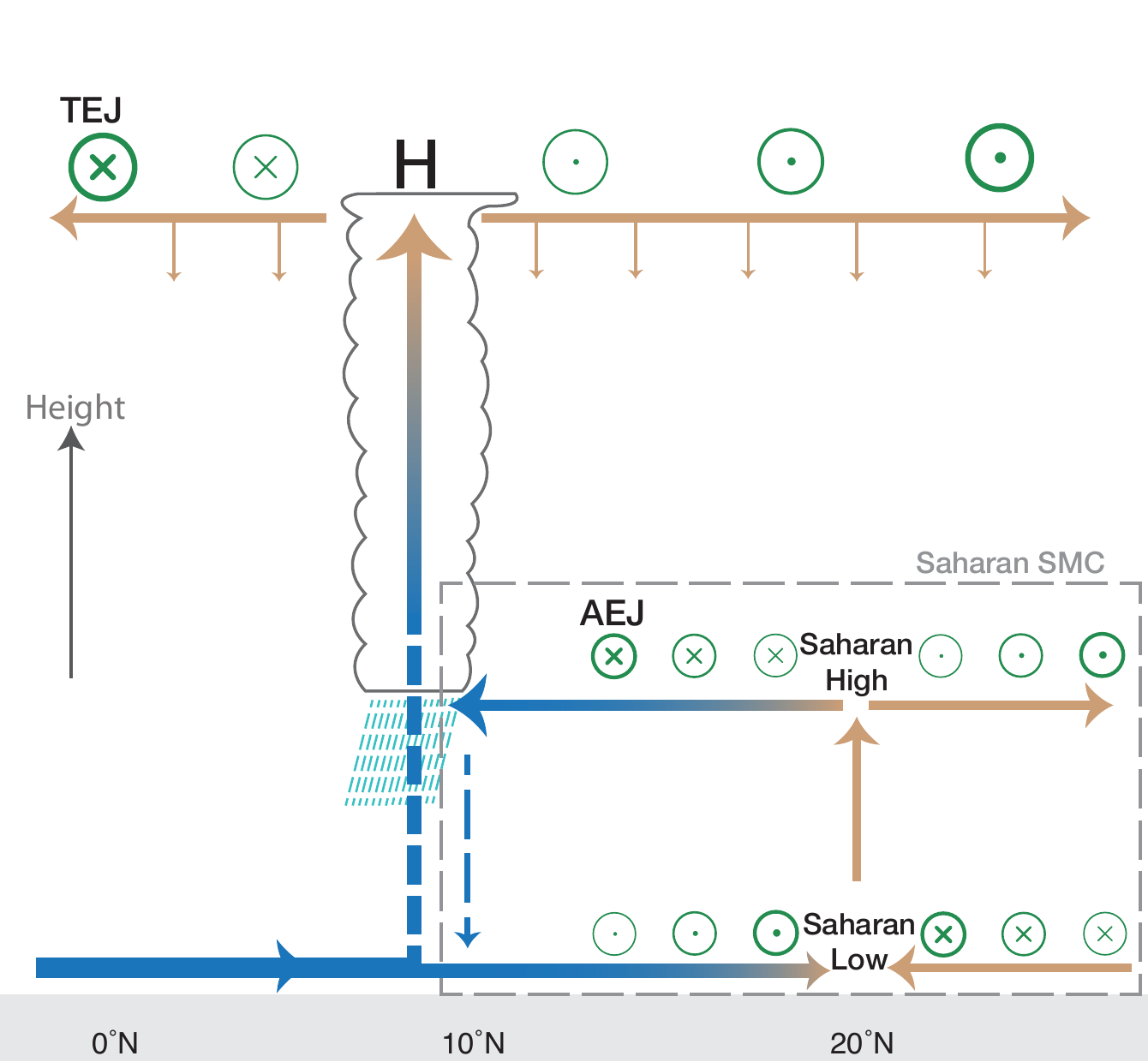}
\caption{The near-surface Saharan Low is indicated at 20\dgN, the mid-level Saharan High anticyclone is indicated by at 20\dgN, and the combination of these is collectively referred to as the Saharan SMC. The mid-level African Easterly Jet (AEJ), upper-level Tropical Easterly Jet (TEJ), and upper level anticyclone in the ITCZ are also shown to help orient the reader. }
\label{fig:schematic}
\end{figure}

The Saharan SMC is related to the Saharan Heat Low, but the two terms describe distinct atmospheric structures.  A heat low or thermal low is a warm region of the atmosphere which has large thickness by the hypsometric relation; these lows are  typically confined below 600 or 700 hPa and are sites of low surface pressure \citep[e.g.][]{Racz1999}.  They include the horizontal balanced flow and the ageostrophic overturning circulations illustrated in Fig.\ \ref{fig:schematic}.  Although the lower troposphere is hot and thus has large thickness across the entire Saharan desert, the term Saharan Heat Low (SHL) is now commonly used to describe the region of maximum low-level atmospheric thickness (LLAT) that exists over the western Sahara \citep[e.g.][]{Lavaysse2009}, and is distinct from the Saharan Low, which is the region of low surface pressure that extends across all of northern Africa \citep[e.g.][]{Biasutti2009}.  The 700 hPa anticyclone in the upper part of the SHL is highly asymmetric, with its southward-flowing eastern branch stretching across the entire eastern half of northern Africa during summer while northward flow in its western branch is weaker and confined to the northwestern African coast \citep{Thorncroft2011}.  The SMC that extends across northern Africa (e.g.\ Fig.\ \ref{fig:smcSectionLon}) should thus not be thought to consist only of ageostrophic flow down the pressure gradient of the SHL, but as a combination of ageostrophic and non-divergent, balanced winds that stretches across all of northern Africa.

In the western Sahara, links between the SHL and central Sahel rainfall have been  studied on a range of time scales. \citet{Lavaysse2009} showed that the SHL exhibits a pronounced seasonal cycle, and moves poleward to its boreal summer position around 20\dg--30\dgN, 7\dgW--5\dgEs just before the onset of the West African monsoon, climatologically in late June. On intraseasonal  and synoptic time scales, strong phases of the SHL are associated with increased moisture transport into the Sahel at low levels, accompanied by increased moisture convergence \citep[e.g.][]{Pu2012} and precipitation over the central and eastern Sahel \citep{Lavaysse2010}. 
\citet{Evan2015} examine the linear trend in the SHL between 1983 and 2009, showing a water vapor-forced warming of the SHL, somewhat consistent with recent work showing amplified warming over the Sahara \citep{Vizy2017}. \citet{Lavaysse2015} show that synoptic and intraseasonal variability of the SHL can influence the seasonal-mean Sahel precipitation, although the mechanisms by which these variations influence rainfall remain unclear.

\begin{figure}
\centering
\includegraphics[width=\halfWidth]{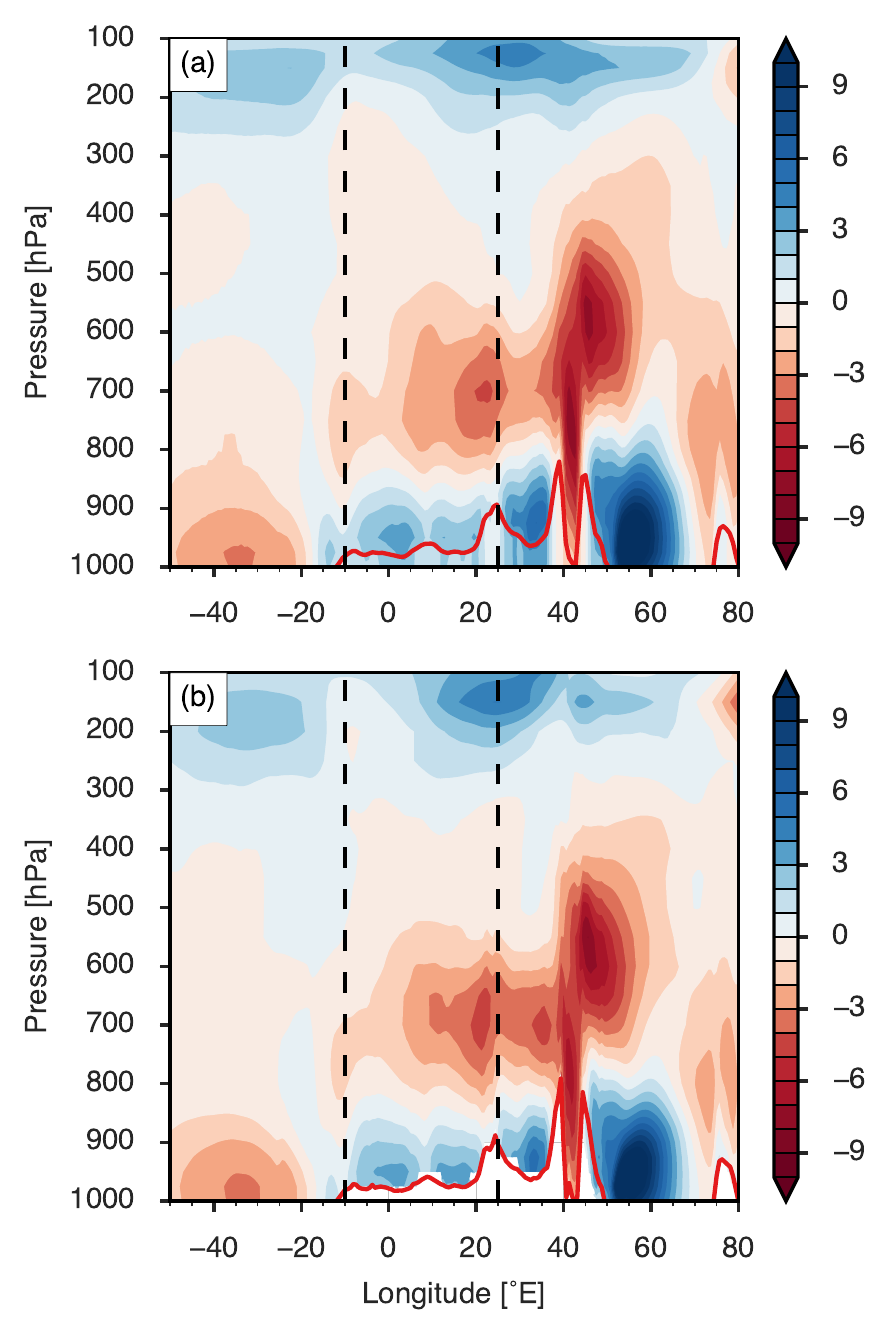}
\caption{JAS climatological meridional wind $v$ at 14\dgNs for (a) ERA-Interim, and (b) MERRA2. 10\dgWs and 25\dgEs boundaries indicated by black dashed lines. Surface pressure indicated by a bold red line. }
\label{fig:smcSectionLon}
\end{figure}

Studies of the SHL in the western Sahara \citep[e.g][]{Lavaysse2010,Lavaysse2010a} and of the Saharan Low that spans all of northern Africa \citep{Haarsma2005,Biasutti2009} suggest that increased Sahel precipitation is caused by a decrease in Saharan surface pressure (or near surface geopotential), which in turn drives an intensification of the mass flux in the Saharan SMC \citep{Martin2014a}. 
However, some studies suggest that the sign of the association between Sahel rainfall and the Saharan SMC should be reversed, with a stronger SMC causing reduced rainfall.  \citet{Peyrille2007} and \citet{Peyrille2007a} used an idealized, zonally symmetric model of the West African monsoon to examine the influence on Sahel rainfall of the large-scale temperature and moisture advection produced by the Saharan SMC. They used an atmospheric reanalysis to show that the Saharan SMC produces near-surface cooling and moistening over the Sahel and Sahara, and warming and drying in the lower mid-troposphere (around 700~hPa). When they separately imposed these low- and mid-level advective tendencies in their idealized model, they found that the low-level cooling and moistening caused increased Sahel rainfall, while the mid-level warming and drying caused decreased Sahel rainfall. The effect of the mid-level warming and drying dominated, so that temperature and moisture advection in the Saharan SMC has a net inhibitory effect on Sahel rainfall.  Furthermore, \citet{Zhang2008} suggested that the transport of hot and dry mid-tropospheric air  into the Sahel precipitation maximum by the Saharan SMC inhibits the northward seasonal migration of the precipitation maximum during early summer.  Mid-tropospheric outflow of hot and dry air from deserts has also been argued to inhibit rainfall in the Indian \citep{Parker2016} and Australian \citep{Xie2010} monsoons.  This is consistent with the demonstrated sensitivity of precipitating convection to drying of the free troposphere above the boundary layer \citep{Derbyshire2004, Holloway2009, Sobel2009}.  In summary, there is evidence for a stronger Saharan SMC having both a positive and negative influence on Sahel rainfall.

To the best of our knowledge, the observed association of interannual variations in Sahel rainfall with the detailed structure of the Saharan SMC has not been examined. It might seem reasonable to assume that the overturning in the Saharan SMC would strengthen as the near-surface Saharan Low strengthens, but given the dominant effect of the mid-level warming and drying on Sahel rainfall suggested by an idealized study \citep{Peyrille2007a}, this would be inconsistent with observations of a strengthening of the near-surface Saharan Low during rainy Sahel years.  Perhaps low-level cooling and moistening by the Saharan SMC has a larger influence on Sahel precipitation in the real world than in the idealized model of \citet{Peyrille2007a}, similar to suggestions for the role of these low-level tendencies in the observed seasonal northward migration of West African rainfall \citep[e.g.][]{Hagos2007, Thorncroft2011, Peyrille2016}.  Or perhaps the Saharan SMC does not strengthen as the near-surface Saharan Low strengthens.  Here we seek to resolve these questions by examining the association of Sahel rainfall with the Saharan SMC at interannual timescales in two atmospheric reanalyses and an idealized model.

The next section of this paper describes our data sources and analysis methods. Section~\ref{sec:results1} discusses the climatology and basic features of the West African monsoon and Saharan SMC. Section~\ref{sec:results2} examines how the horizontal structure of the near-surface Saharan Low and the Saharan High covary with Sahel precipitation, and is followed by a section detailing the vertical structure of the circulation changes, with emphasis on the divergent component of the flow. Section~\ref{sec:results5} compares all of these observationally based results with output from an idealized $\beta$-plane model. We close with a discussion of implications and caveats in section~\ref{sec:discussion}.

\section{Methods}
\label{sec:methods}
We obtain winds, geopotential height, temperature, and humidity from the ERA-Interim reanalysis  \citep{Dee2011}, which is produced by the European Centre for Medium-Range Weather Forecasts (ECMWF) and is used here for 1979-2015. ERA-Interim is a third generation reanalysis with data assimilation based on 12-hourly four dimensional variational analysis (4D-Var). The dynamics are calculated on a T255 (approximately 80 km) global grid, with 60 vertical levels from the surface to 0.1~hPa. We also use NASA's Modern-Era Retrospective analysis for Research and Applications, Version 2 \citep[MERRA2;][]{Gelaro2016}, which is a third generation reanalysis produced on a 0.5\dg$\times$0.625\dgs \, cubed-sphere grid with 72 vertical levels from the surface to 0.1~hPa. MERRA2 is not available for 1979, so here we use 1980-2015. All climatologies and regressions shown here use ERA-Interim data unless MERRA2 is explicitly indicated. MERRA2 versions of all applicable figures are provided in supplemental materials.

The Global Precipitation Climatology Project (GPCP) dataset \citep{Adler2003}, which is a combination of land rain gauge and satellite-based precipitation measurements, is used as the primary precipitation dataset for this study. The Global Precipitation Climatology Centre \citep[GPCC,][]{Schneider2014} dataset, based on corrected gridded rain gauge data, was also examined and found to produce no substantial changes in our conclusions. All data was obtained at monthly mean resolution.

\begin{figure*}
\centering
\includegraphics[width=\fullWidth]{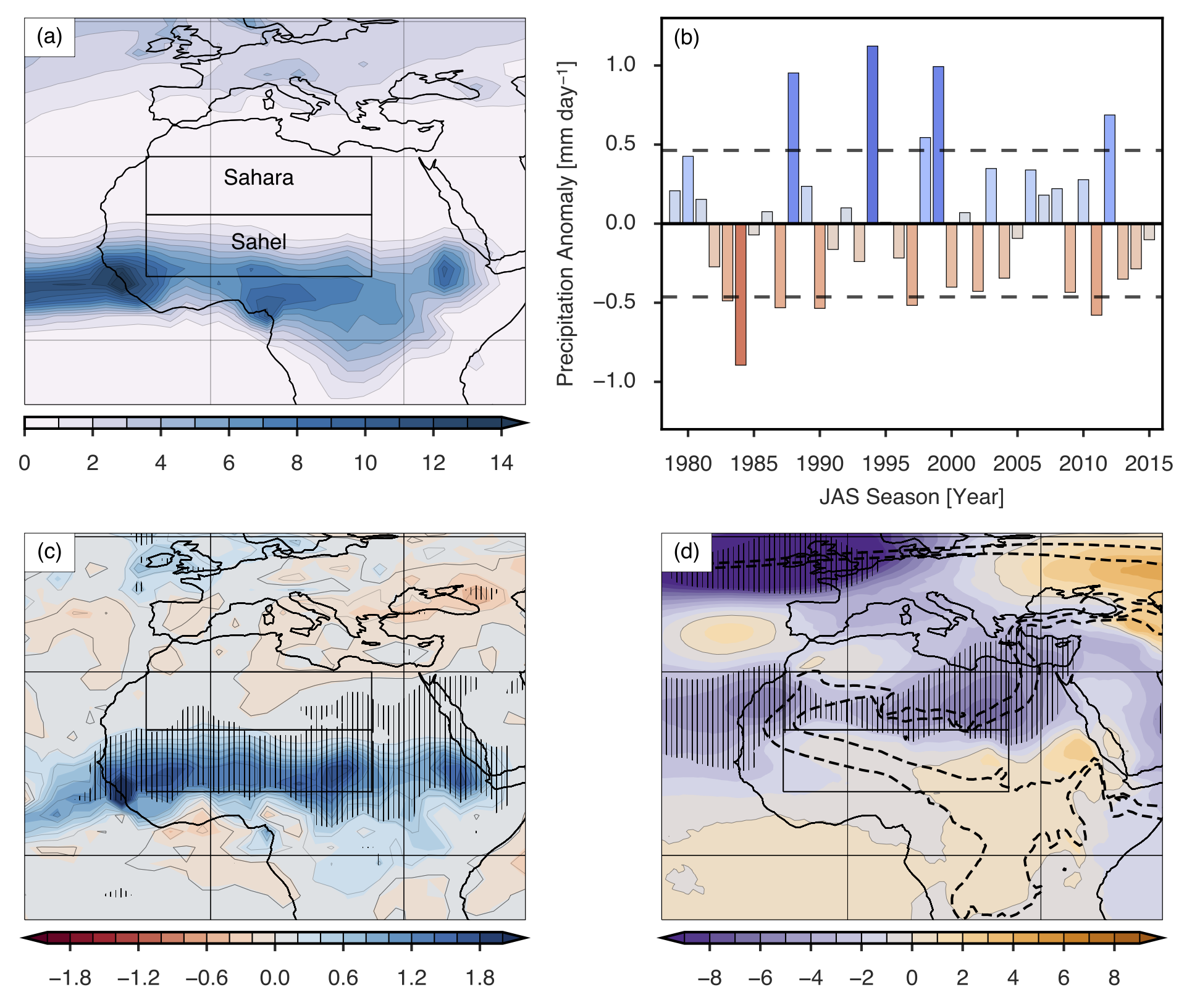}
\caption{(a) June-September precipitation climatology based on GPCP data, with Sahel and Sahara boxes indicated. Units mm day\mones (b) Detrended interannual time series of anomalous precipitation over the Sahel box derived from GPCP data, with $\pm$1 standard deviation bounds are indicated by dashed lines. (c) Regression slope of local detrended GPCP precipitation onto Sahel area averaged precipitation in mm~day\mones per mm~day\mone. (d) ERA-Interim regression slope of detrended Z925 (m~mm\mone~day; colors) and climatological $\Delta$Z925 (dashed contours mark $-10$~m and $-20$~m isolines).  Statistically significant regression slopes ($p<0.05$) are hatched.}
\label{fig:overview}
\end{figure*}

We take seasonal averages over the July-September (JAS) season for all variables, as this is the peak of the Sahelian phase \citep[e.g.][]{Nicholson2013} of the African monsoon. We conducted a sensitivity study to alternate choices of season, specifically JJAS and JA, and found very few differences in our results.  The average over all JAS periods in the reanalysis time period is referred to as the climatology. Fig.~\ref{fig:overview}a shows the GPCP JAS climatology of precipitation, with the Sahelian phase rainfall peak near 10\dgN. 

We are interested in the Saharan SMC (Fig.~\ref{fig:smcSectionLon}), so we choose  10\dgW--25\dgEs as zonal bounds for our analyses. These zonal bounds  encompass most of northern hemisphere Africa but exclude coasts, the East African highlands, and the Arabian Desert. In the 10\dgW--25\dgEs region, the latitudes of 10\dgNs to 20\dgNs are defined as the Sahel, and 20\dgNs to 30\dgNs as the Sahara, with both regions delineated with boxes in ~\rFigOverview{a}. Anomalies of precipitation were calculated with respect to the JAS climatology, averaging over the Sahel box and subtracting the linear trend ($0.26\pm0.15$~mm~day\mone~decade\mone) and the climatological mean (3.31~mm~day\mone).  The resulting timeseries is hereafter referred to as `Sahel precipitation' (Fig.~\ref{fig:overview}b) and forms the basis of most regressions presented here. We also removed the linear trend from all other fields regressed against this Sahel precipitation time series. Over the reanalysis period of 1979-2015, Sahel precipitation lacks statistically significant interannual autocorrelations (not shown), distinguishing variability in this more recent period from the persistent interdecadal droughts that characterized parts of the twentieth century and that were largely attributed to global variations in SST \citep{Giannini2003}. \citet{Losada2012} noted the non-stationarity of the relationship between Sahel precipitation and SST in the different ocean basins over the twentieth century, and showed a marked transition in SST dependence in the 1970s, with a largely stationary regime of SST dependence since then. In our post-1970s period, precipitation mostly exhibits a ``monopole'' spatial pattern (\rFigOverview{c}), with single-signed precipitation anomalies extending from the Gulf of Guinea across the Sahel. This is in contrast to meridional ``dipole'' patterns of precipitation anomalies observed during the 1920s-1970s \citep[Fig.~14 of][]{Nicholson2013}. This has implications for the generality of our results, which we discuss further in section~\ref{sec:discussion}. 

As many studies on the SHL focus on the 10\dgW--10\dgEs region, we examined the sensitivity of our results to the zonal bounds, with precipitation timeseries created by averaging over the western portion (10\dgW--10\dgE) and eastern parts (10\dgE--25\dgE) of the Sahel. These two timeseries were highly correlated at interannual time scales, with a coefficient of determination of $R^2=0.72$. Furthermore, both eastern and western Sahel precipitation timeseries were individually very highly correlated ($R^2 > 0.9$) with the  precipitation timeseries presented in Fig.~\ref{fig:overview}b. The entire analysis presented here was repeated for the eastern and western portions of the domain, with the appropriate eastern or western precipitation timeseries. The appendix explores some modest differences between these eastern and western subdomains, but our general conclusions also hold in these subdomains. This lack of sensitivity to the zonal bounds is consistent with the Saharan SMC (Fig.\ \ref{fig:smcSectionLon} and interannual variations in Sahel rainfall (Fig.\ \ref{fig:overview}c) being coherent across the entire zonal extent of Africa. 
\citet{Lavaysse2015} regress GPCP precipitation on a metric of SHL strength, in contrast to our method of regressing precipitation onto the area-averaged Sahel precipitation, and also obtain coherent rainfall anomalies across the entire Sahel (10\dgN--20\dgNs, 10\dgW--25\dgE) at certain timescales (see their Fig.~11d). 

For the idealized model portion of this study, we analyzed the same integrations presented in \citet{Shekhar2016}. These used the Weather Research and Forecasting (WRF) model, version 3.3 \citep{Skamarock2008}, modified to run on an equatorial $\beta$-plane in a meridional channel at 15~km resolution with 41 vertical levels.  The domain was 20\dg$\times$140\dgs\ in the zonal and meridional directions, respectively, with periodic boundary conditions in the zonal direction and closed boundary conditions in the meridional direction. A continent was prescribed from 5\dgNs to 32\dgN, divided into a grassland from 5\dgN--12\dgNs and a desert from 12\dgN--32\dgN, with interactive surface temperature but prescribed soil moisture and other properties from the WRF land surface database. The remainder of the domain was ocean with a prescribed, idealized SST distribution representative of that observed during boreal summer near Africa. Perpetual July 15 insolation was imposed, with the diurnal cycle retained. As shown by \citet{Xie1999}, fixed SST, prescribed soil moisture, and perpetual summer form a physically consistent set of simplifications, and allowing any one be more realistic requires allowing all three to be more realistic to avoid unphysical variability, which would greatly increase computational requirements.  A total of thirteen model integrations were performed for one model year after a three month spinup, yielding the same amount of output for each integration as four three-month summer seasons. One integration was chosen as the control, and others were forced by modifications of the specified desert surface albedo, the prescribed SST, or both. These form an ensemble of integrations in which the monsoon precipitation varies in response to the SST and surface albedo forcings.  These integrations were documented more thoroughly in \citet{Shekhar2016}, where they were used to examine energy-based diagnostics of ITCZ location.

Statistical analyses were performed using Python programming language packages  Iris \citep{Iris2016}, Seaborn \citep{Seaborn2016}, and Statsmodels \citep{statsmodels2010}. For linear regressions, we test for a nonzero slope using a two-sided Student's $t$ test at the $p<0.05$ level. For some linear regressions, we also obtain a 95\% confidence interval for the slope using a bootstrapping technique on the joint probability distribution of slope and intercept, as given in the Seaborn package. We tested the sensitivity of linear regressions to outliers using the robust regression \citep[e.g.][]{Rousseeuw2005} feature of the Statsmodels package, and although certain confidence intervals narrow and shift slightly, no substantial qualitative differences were obtained. We also examined the possibility of using nonlinear (quadratic and cubic) models for regression, and found a statistical preference for nonlinear models for less than 4\% of gridpoints at the 95\% confidence level, an amount attributable to chance. 

\section{Basic features of the Saharan Low and SMC}
\label{sec:results1}

We begin by examining the association of interannual variations in the Saharan Low with those of Sahel rainfall, then detailing the climatological mean structure of the Saharan SMC.  Although these analyses resemble those published previously \citep[e.g.][]{Biasutti2009, Thorncroft2011}, the results confirm that the main features of interest are found in the ERA-Interim and MERRA2 reanalyses and serve as a necessary reference for the rest of this paper.

\begin{figure*}
\centering
\includegraphics[width=\fullWidth]{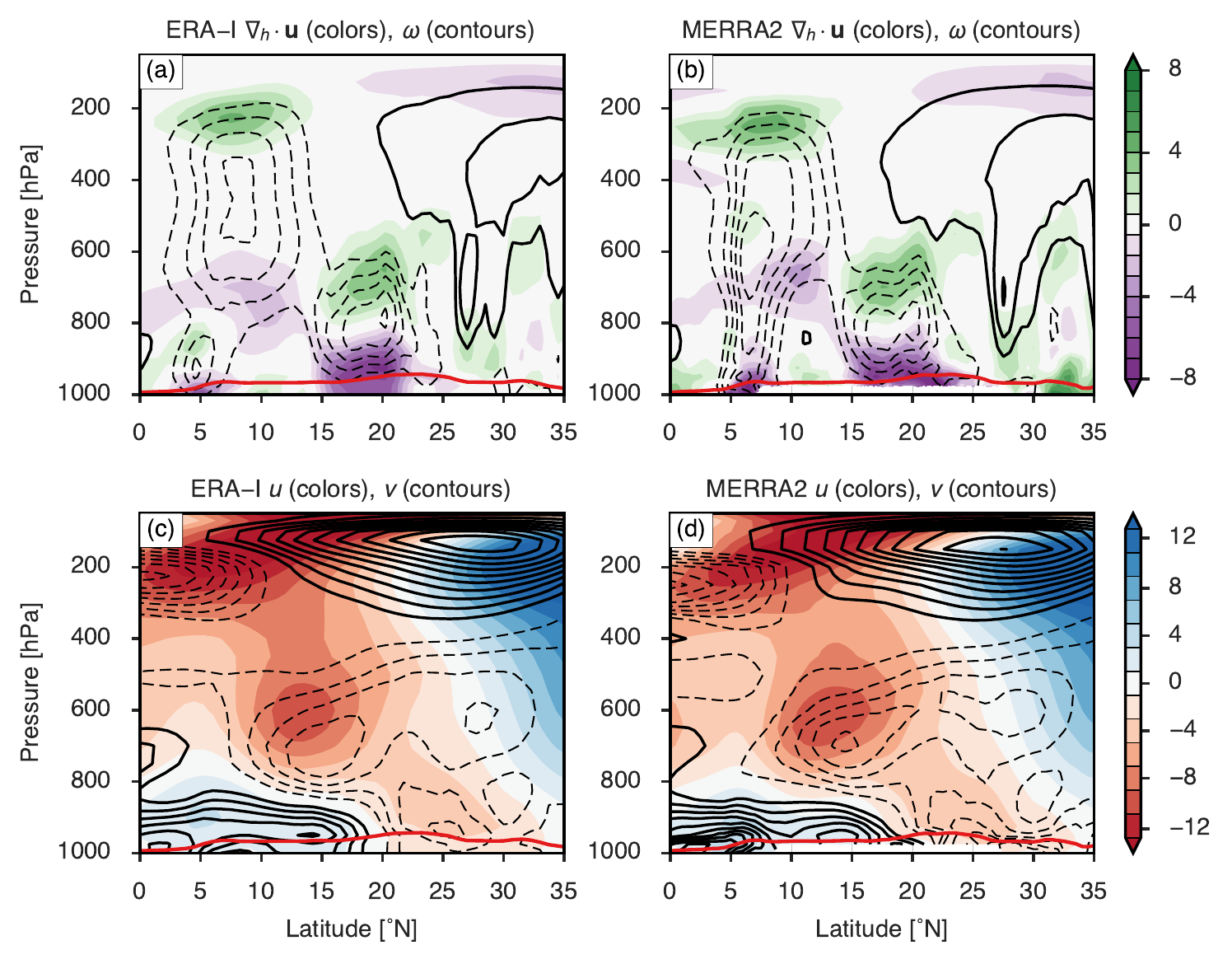}
\caption{Climatological JAS zonal mean (a, b) Divergence (colors; 10$^{-6}$~s\mone) and pressure velocity $\omega$ (contours every $0.5$~hPa~hr\mone). (c, d) Zonal wind $u$ (colors) and meridional wind $v$ (contours every m~s\mone). ERA-Interim used for panels a and c, whereas MERRA2 is used for panels b and d. All panels have zero contours omitted, and negative contours are dashed. } 
\label{fig:shlClim}
\end{figure*}

\citet{Biasutti2009} obtained a correlation of high Sahel rainfall and negative anomalies of 925~hPa geopotential height (Z925) over the Sahara desert. We reproduce this correlation in reanalyses (Fig.~\ref{fig:overview}d). A regression of mean sea level pressure on Sahel rainfall \citep[as in][]{Haarsma2005} yields a similar pattern (not shown).  These previous studies interpreted these patterns of Z925 and mean sea level pressure as indicative of a strengthening of the Saharan Low.  But the near-surface Saharan Low stretches zonally across northern Africa and is centered around 20\dgNs while the decrease in Z925 is confined to the northern and western sides of that climatological trough.  Since a strengthening of the Saharan Low would consist of negative anomalies of Z925 centered over the climatological minimum Z925, the concentration of negative anomalies northward and westward of that climatological minimum indicates that the Saharan Low is expanding northward and westward rather than simply strengthening. Much of the northward expansion occurs over northeastern Africa, consistent with the patterns in anomalous Z925 and mean sea level pressure seen in \citet{Biasutti2009} and \citet{Haarsma2005}.

The near-surface Saharan Low is part of the three-dimensional Saharan SMC, as mentioned in the introduction. The ascending branch of the Saharan SMC is strongest at 20\dgNs in the climatological and zonal mean (\rFigSHLClim{a,b}). An examination of the zonal structure in the appendix shows the latitude of peak ascent moves equatorward as one moves east, from 23\dgNs in the west to 19\dgNs in the east. Low level mass convergence occurs between the surface and 800~hPa, with peak convergence at 925~hPa in the near-surface Saharan Low.  Divergence occurs in the 800--550~hPa layer, with peak divergence at 700~hPa associated with the Saharan High. The 925~hPa and 700~hPa levels are taken as representative of the near-surface Saharan Low and the Saharan High, respectively, in the next two sections. The maximum deep ascent in the ITCZ is located much further south, around 8\dgNs in ERA-Interim (\rFigSHLClim{a}) and 6\dgNs in MERRA2 (\rFigSHLClim{b}). Vertical structures of vertical velocity are somewhat different between the two reanalyses, perhaps because of differences in convective parametrizations. There is weak time-mean divergence and subsidence in the near-surface layer around 10\dgNs in the precipitating region, a likely signal of strong time-mean subsidence and sporadic ascent due to precipitation.  

The near-surface zonal and meridional winds change sign around 20\dgNs in the climatological zonal mean (\rFigSHLClim{c,d}). This change in sign, a feature sometimes described as the ITD (intertropical discontinuity), is not entirely zonally symmetric, and shifts equatorward by a few degrees latitude as one moves eastward within our analysis domain (see section~\ref{sec:results2}). Around 700~hPa, there is a peak in the equatorward flow as air travels in the time-mean Saharan SMC toward the ITCZ.  The African Easterly Jet \citep[AEJ; e.g.][]{Thorncroft1999} exists in thermal wind balance around 600~hPa and 14\dgN. At upper levels, the tropical easterly jet, meridional flow in the upper branch of the Hadley circulation, and the midlatitude jet stream are also visible. In the next two sections, we examine how the strength and spatial structure of the Saharan SMC covary with Sahel precipitation.

\section{Horizontal structure of Saharan SMC changes}
\label{sec:results2}

\begin{figure*}
\centering
\includegraphics[width=34pc]{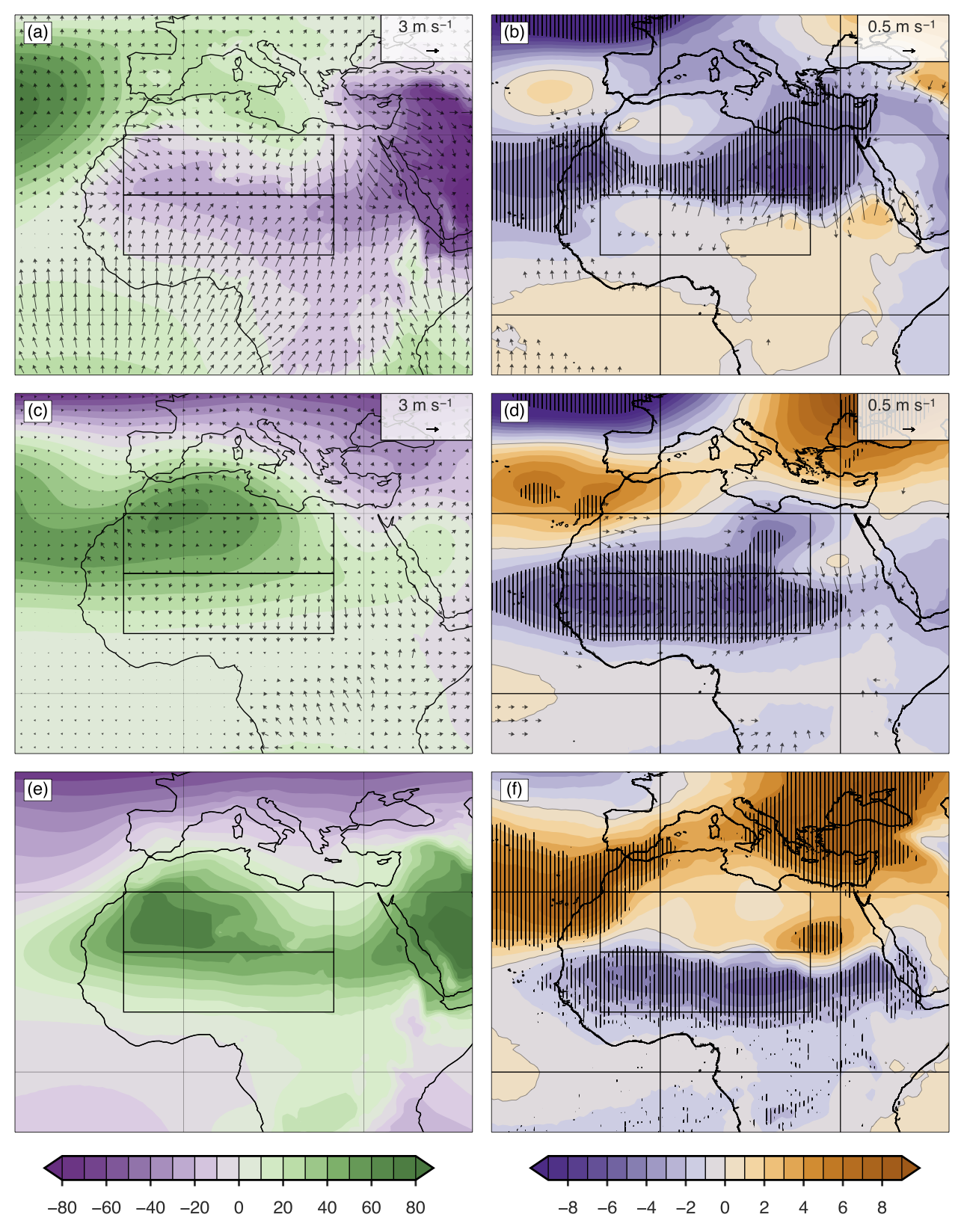}
\caption{All panels use ERA-Interim data. (a, c, e) Colors show climatological JAS Z925, Z700, and LLAT respectively in units of m. Constant values of 790~m, 3150~m, and 2260~m respectively were subtracted to reduce them to the same color scale. Arrows show the divergent component of the wind $\mathbf{u}_\chi$ for the same 925 and 700~hPa levels respectively. (b, d, f) Colors show the regression slope of detrended Z925, Z700, and LLAT onto GPCP Sahel precipitation respectively (m~mm\mone~day), with hatched regions indicating statistical significance. Divergent wind regression slope for same 925 and 700~hPa levels respectively shown in arrows, with only regions where $u_\chi$ or $v_\chi$ is statistically significant and $|\mathbf{u}_\chi| > 0.1$~m~s\mone~mm\mone~day are drawn. }
\label{fig:ghtMaps}
\end{figure*}

\begin{figure*}
\centering
\includegraphics[width=\fullWidth]{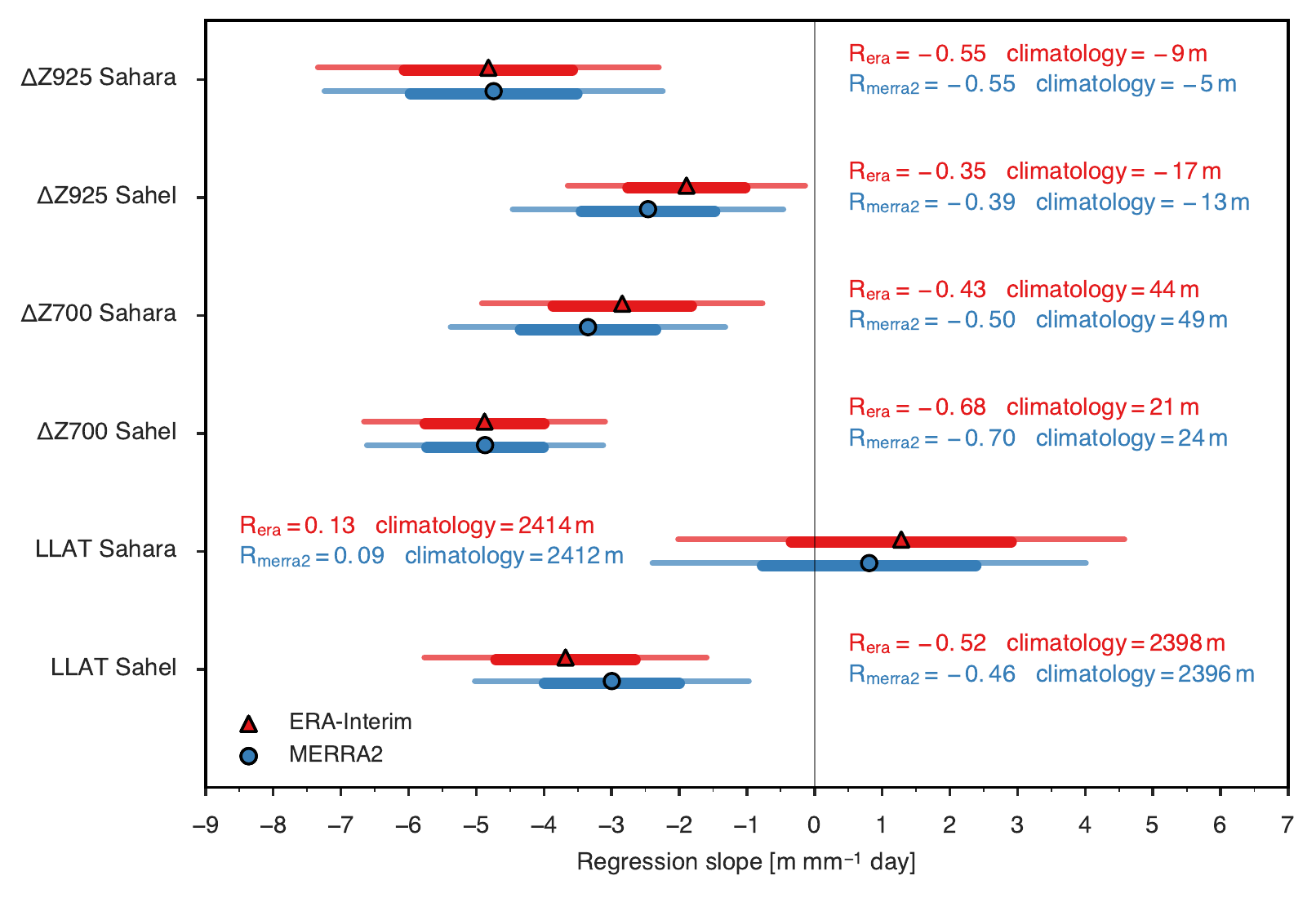}
\caption{$\Delta$Z925, $\Delta$Z700, and LLAT were averaged over the Sahara and Sahel in both ERA-Interim and MERRA2 reanalyses, detrended, and regressed against Sahel precipitation. 68\% (thick) and 95\% (thin) confidence intervals of the regression slope are shown. Climatological values of each quantity and the Pearson $R$ coefficient of the regression are indicated on the figure. Critical R values with $n-2=34$ degrees of freedom are 0.329, 0.423, and 0.525 at the 0.05, 0.01, and 0.001 levels.}
\label{fig:ghtCI}
\end{figure*}

Since the geopotential height and divergent wind together provide a nearly complete depiction of the horizontal circulation, we start by examining the horizontal structure of geopotential  and divergent wind variations at 925 and 700~hPa.  \rFigGhtMaps{a} shows the 925~hPa climatological trough extending across northern Africa during JAS around 20\dgN. The previously discussed ITD exists at the center of this trough, which shifts slightly toward the equator in the eastern portion of the domain.  Geostrophic flow moves cyclonically around the trough (not shown). Divergent winds converge into the trough, and cross-equatorial southerly flow in the low-level branch of the Hadley cell is also visible over the Gulf of Guinea. When regressed on Sahel precipitation, we see a spatially heterogeneous but nearly single-signed decrease in Z925 (\rFigGhtMaps{b}) north of 20\dgNs over northern Africa and portions of the Atlantic (the dynamical implications of this decrease are discussed in the next two sections). Although changes in Z925 are not statistically significant over the Sahel, there are weak anomalies of northerly divergent flow in the southern Sahel during wet years that indicate a weakening of the ageostrophic northward monsoon flow at 925 hPa.  At the northern boundary of the Sahel, anomalous southerly wind flows across the climatological ITD into the region of anomalously low Z925. Effectively, this moves the ITD and minimum of the Saharan Low poleward during wet years (a quantitative measure of the meridional shift in the Saharan Low is provided in the next section).

\rFigGhtMaps{c} shows the horizontal structure of the climatological mean Saharan High, with gradients in Z700 in geostrophic balance with an anticyclone over most of the Saharan region.  There is a substantial zonal gradient in Z700 over the central and eastern Sahara, and a sharp meridional gradient over the Sahel in balance with the AEJ. The peak in climatological Z700 lies poleward and westward of the trough in climatological Z925. Downgradient divergent flow occurs south and northwest of the 700~hPa high, constituting the divergent northerly outflow in the upper branch of the Saharan SMC. The regression of Z700 onto Sahel precipitation (\rFigGhtMaps{d}) shows a statistically significant decrease centered in the Sahel implying a strong anomalous cyclonic circulation during wet years.  This mid-level anomalous cyclone is also visible at 600~hPa (not shown), and is well supported by observational evidence of a northward shift and weakening in the AEJ during wet Sahel years \citep[e.g.][]{Nicholson2013}.  At 700~hPa, the anomalous divergent wind converges into the center of the anomalous low, around 15\dgNs in the eastern Sahel.

We expect the 925 and 700~hPa surfaces to be affected differently by a strengthening of the shallow SMC compared to a strengthening of the deep, precipitating monsoon circulation.  A stronger shallow SMC is expected to be accompanied by a reduction in Z925 in the Saharan Low, an increase in Z700 in the Saharan High, and an intensification of the divergent, overturning circulation that flows down the geopotential gradients at these two levels.  In contrast, enhanced precipitation is expected to be accompanied by enhanced ascent in a deep circulation that can be approximated by a first-baroclinic mode; strengthening of such a first-baroclinic mode will include decreases in geopotential height in the entire lower and middle troposphere and increases in geopotential in the upper troposphere [see \citet{Neelin2000} for a derivation of the structure of a typical tropical first-baroclinic mode, and \citet{Zhang2008} or \citet{Nie2010} for illustration of how the shallow Saharan SMC coexists with a precipitating first-baroclinic mode structure over West Africa].  The 925~hPa and 700~hPa surfaces are thus expected to have opposite vertical displacements as the Saharan SMC intensifies, simply because the Saharan LLAT would need to increase to maintain a stronger ageostrophic overturning, absent any large changes in drag.  In contrast, those two surfaces are both expected to move downward as the deep, precipitating circulation strengthens.  \rFigGhtMaps{} (panels b and d) shows that the 700 and 925~hPa surfaces both move downward during wet Sahel years, providing no evidence for a strengthening of the shallow SMC.  Furthermore, when geopotential heights are averaged over our Sahel and Sahara boxes, statistically significant negative anomalies in Z925 and Z700 are found during wet years in both regions (\rFigGhtCI{}).  Thus, geopotential variations at 925 and 700~hPa are inconsistent with the hypothesis that the Saharan SMC strengthens during wet Sahel years.

These changes in structure can also be viewed in terms of the thickness of the lower troposphere, but one must remember that LLAT will increase during a strengthening of the shallow heat low circulation and during a strengthening of the deep precipitating circulation.  Shifts in the midlatitude barotropic flow (e.g.\ the jet stream) can also project on LLAT.  The LLAT climatology (\rFigGhtMaps{e}) shows a maximum over the western Sahara, with relatively high LLAT extending east along the the 20\dgNs parallel, approximately following the ITD.  Thus, although the Saharan Heat Low is commonly thought of as being confined to the western Sahara \citep[e.g.][]{Lavaysse2009}, a band of high LLAT extends eastward across the entire Sahara into the Arabian desert heat low.  

The regression of LLAT onto Sahel precipitation (\rFigGhtMaps{f}) shows increases in LLAT centered to the northeast and northwest of our Saharan box, but the regression slope is not statistically distinct from zero over most of the Sahara. Over nearly all of the Sahel, a statistically significant decrease in the LLAT is apparent.  The weak zonal variation in the anomalies shown in the right column of \rFigGhtMaps{} is important:  there is no thickening of the lower troposphere in the Western Sahara, where the LLAT is climatologically highest. If the Saharan Heat Low or the Saharan SMC were strengthening, we would see increased LLAT at the climatological maximum LLAT. The substantial increases in LLAT over eastern Europe, the Mediterranean, and the Atlantic might indicate interactions with the midlatitudes, perhaps through mechanisms proposed by \citet{Vizy2009} and \citet{Lavaysse2010a}.  In summary, there is a decrease in LLAT on the equatorward side of the maximum LLAT, and the strong increases in LLAT are centered off the coasts of Africa, poleward of the LLAT maximum. The changes in LLAT thus seem more consistent with a poleward shift, rather than a strengthening, of the thermal low and the Saharan SMC  during wet years. These results do not change appreciably if an alternate upper bound (e.g. 600 or 500~hPa) is used to define the LLAT, which would allow us to capture a deepening of the heat low \citep[e.g.][]{Evan2015}. 

We have thus far examined variations in geopotential height, but it is horizontal gradients in geopotential that are dynamically relevant.  Changes in these gradients can be assessed by eye in maps of the horizontal distribution of geopotential, but we now wish to horizontally average the changes in geopotential over our analysis regions.  So for every JAS season we subtract the tropical-mean (23\dgS--23\dgN) geopotential height from the actual height at each pressure-level:  $\Delta\mathrm{Z}=\mathrm{Z}-[\mathrm{Z}]_\mathrm{global\ tropics}$.  This methodology, which was also performed by \citet{Biasutti2009}, eliminates dynamically inconsequential changes in geopotential associated with tropical-mean warming or cooling due to, e.g., ENSO.

\begin{figure*}
\centering
\includegraphics[width=\fullWidth]{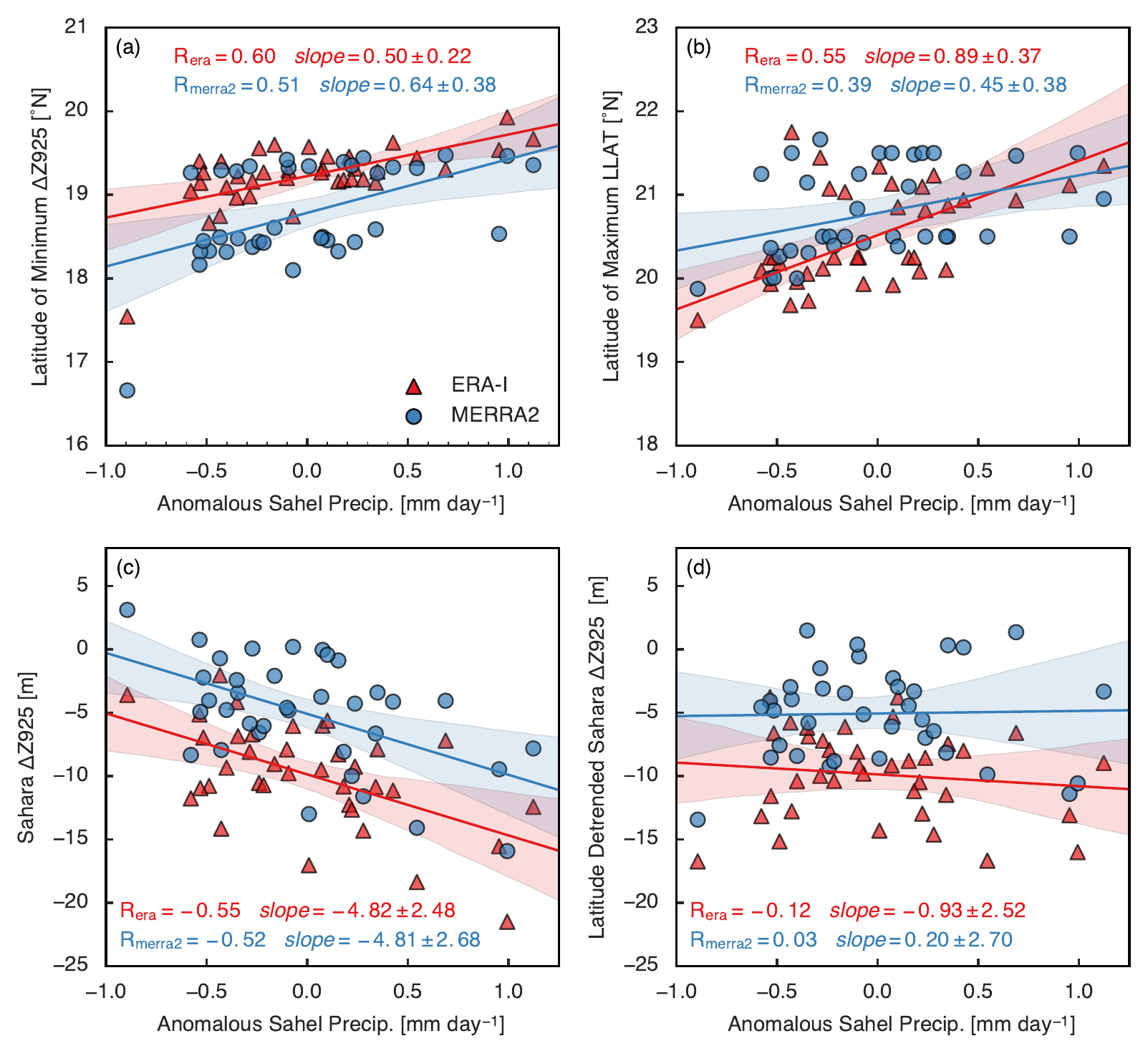}
\caption{Regressions against Sahel precipitation of: (a) 925~hPa trough location (latitude of minimum 925~hPa $\Delta$Z925), (b) Latitude of maximum LLAT, (c) Sahara $\Delta$Z925, and (d) Sahara $\Delta$Z925 with linear dependence on heat low trough location removed. All quantities had the long term time trend removed after area averaging. }
\label{fig:hlLocation}
\end{figure*}

Regressions of detrended $\Delta$Z on Sahel precipitation quantitatively reproduce the result, shown by \citet{Biasutti2009}, of decreased Saharan $\Delta$Z925 during wet Sahel years (\rFigGhtCI{}).  The ERA-Interim results are quantitatively indistinguishable from the MERRA2 results.  We also see decreased $\Delta$Z925 over the Sahel, and an inspection of the climatological values (text insets in \rFigGhtCI{}) indicates that the meridional gradient of $\Delta$Z925 between the Sahara and Sahel flattens during wet years. The Sahel LLAT decreases as a consequence of $\Delta$Z925 decreasing less than $\Delta$Z700 during wet years (in this case, removal of the tropical mean has nearly zero effect, with variations in $\Delta\mathrm{Z700}-\Delta\mathrm{Z925}$ being nearly equal to variations in $\mathrm{Z700-Z925}$).  This is inconsistent with the idea that a classic first-baroclinic mode structure intensifies over the Sahel, and we will show in the next section that the vertical profile of the anomalous convergence during wet years also differs from that of a classic first-baroclinic mode, but still provides no evidence for a strengthening of the Saharan low during wet years.  Changes in Saharan LLAT are not statistically distinguishable from zero.  This analysis confirms that, even when data is horizontally averaged, no strengthening of the Saharan SMC can be detected in the LLAT.

Since we found some evidence of a northward expansion of the Z925 trough into the Sahara during wet Sahel years (\rFigGhtMaps{b}), we ask how much of the change geopotential can be attributed to a simple meridional shift of the trough.  We define the ``trough latitude'' by taking a 10\dgW--25\dgEs zonal average of $\Delta$Z925 then finding the latitude of the minimum, using cubic splines to interpolate between reanalysis gridpoints. The trough latitude  exhibits a strong positive correlation with Sahel rainfall (\rFigHLLocation{a}), with a slope of $0.50\pm0.22$~degrees~mm\mone~day in ERA-Interim. Both reanalyses contain an influential data point in 1984, the driest year in the reanalysis period, with the trough latitude about 1 degree farther equatorward in that year than in all other years. Removing this extreme data point or using robust linear regression decreases slopes to approximately 0.45 degrees mm\mone day, but does not qualitatively change the relationship between the trough latitude and Sahel precipitation. The MERRA2 data exhibit a somewhat bimodal distribution in trough latitude for which we do not have an explanation. 

The trough latitude might be influenced by changes in both the thermal low and the deep, precipitating circulation, so we also examine the latitude of the maximum zonally averaged LLAT as a more direct measure of the position of the thermal low (\rFigHLLocation{b}).  The latitude of maximum LLAT typically lies a few degrees poleward of the trough latitude and also has higher interannual variability.  Nevertheless, it shows a northward shift of the zonally averaged thermal low during wet Sahel years. The regression slopes for the 925 hPa trough latitude and the latitude of maximum LLAT are statistically indistinguishable (i.e.\ their confidence intervals overlap).  

We now return to the question of how much of the drop in Saharan $\Delta$Z925 can be attributed to a poleward shift of the trough into the Sahara, and answer this question by statistically removing the effect of this shift from the Saharan $\Delta$Z925. 
The regression of Saharan $\Delta$Z925 on trough latitude produces a slope of $-4.8\pm2.5$ meters per degree latitude (\rFigHLLocation{c}).  When this dependence is subtracted from the Saharan-averaged $\Delta$Z925 to create a ``latitude-detrended $\Delta$Z925'', the resulting quantity has no statistically significant relationship with Sahel precipitation (\rFigHLLocation{d}). This suggests that previous findings of changes in the Saharan Low during wet Sahel years \citep[e.g.][]{Haarsma2005, Biasutti2009} should be interpreted as a shift of the low-level trough into the Sahara.  As the next section will show, many dynamic and thermodynamic quantities also show a poleward shift.  

\section{Vertical structure of changes in the Saharan SMC}
\label{sec:results3}

\begin{figure*}
\centering
\includegraphics[width=38pc]{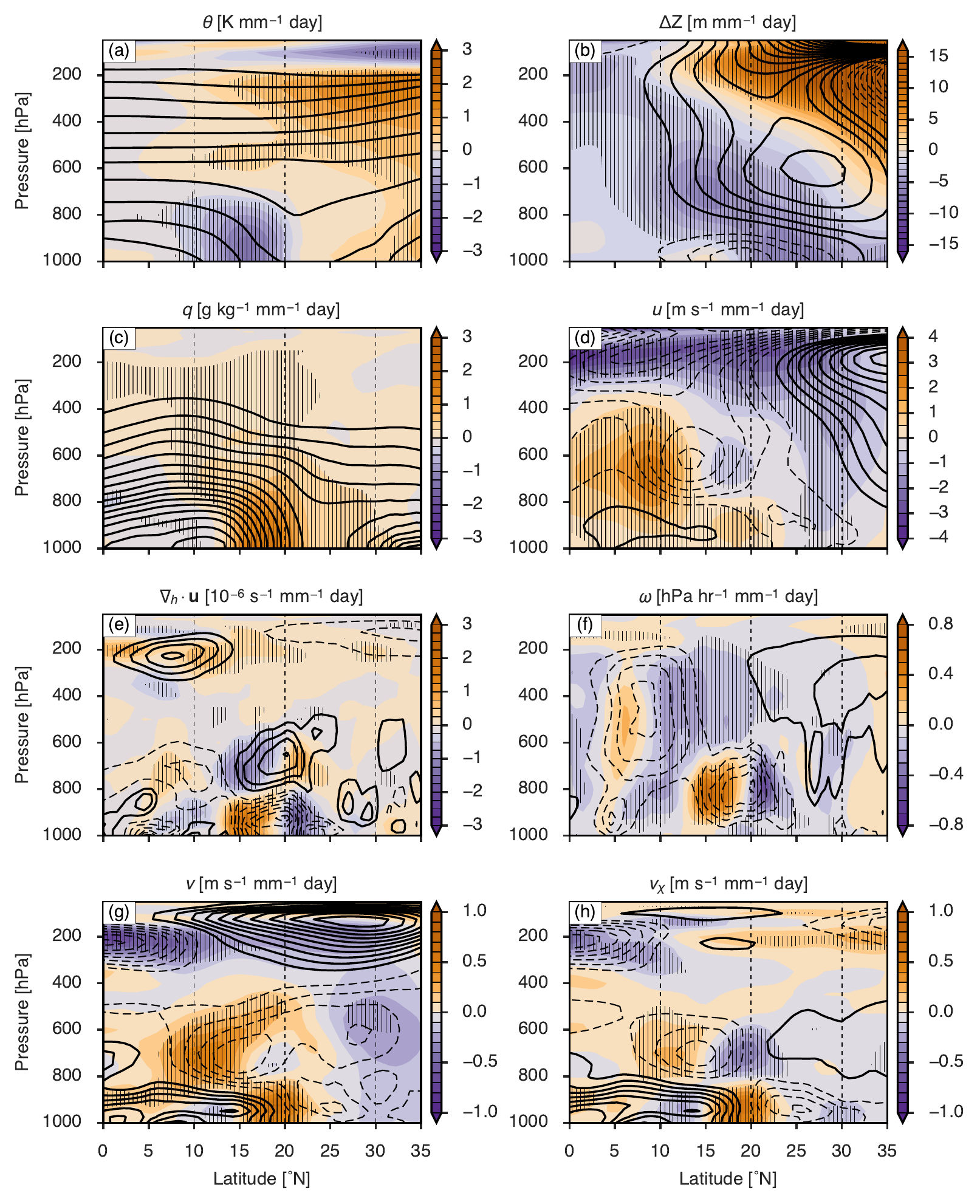}
\caption{All panels show ERA-Interim zonal averages over 10\dgW --25\dgEs. Colors indicate regression slope of the detrended quantity onto Sahel precipitation, hatching indicates statistical significance, and black contours indicate climatology (negative dashed, zero omitted). Units on regression slope imply per mm~day\mone. 
(a) $\theta$ [K; contours every 5].
(b) $\Delta$Z [meters; contours every 10]. 
(c) specific humidity [g~kg\mone; contours every 1]
(d) $u$ [m~s\mone; contours every 2].
(e) Horizontal divergence [10$^{-6}$~s\mone; contours every 1].
(f) $\omega$ [hPa~hr\mone; contours every 0.5].
(g) $v$ [m~s\mone; contours every 0.5].
(h) $v_\chi$ [m~s\mone; contours every 0.5].
}
\label{fig:vertSections}
\end{figure*}

We now examine the vertical structure of interannual variations in the West African monsoon circulation, with focus on the Saharan SMC.  We first briefly examine the poleward shift of the SMC during wet monsoon years, then discuss variations in the strength of the divergent component of the circulation. While a few of the vertical structures shown here were presented in previous work \citep[e.g.][]{Nicholson2001, Nie2010, Lavaysse2010}, they are included to provide a complete picture of the dynamical changes that constitute a weakening and poleward shift of the Saharan SMC.

The climatological low-level potential temperature is maximum around 22\dgNs (\rFigVertSections{a}), poleward of the ITD. As Sahel precipitation increases, there is substantial cooling over the Sahel below 700~hPa, likely due to evaporative cooling of the land surface and reduced surface sensible heat fluxes into the boundary layer.  Above 700~hPa, warming occurs poleward of 10\dgNs into the midlatitudes. By the hypsometric relation, this has a direct impact on the the $\Delta$Z climatology (\rFigVertSections{b}), which shows the near surface Saharan Low centered around 18\dgNs in the climatology and the mid-tropospheric Saharan High centered around 25\dgN. The regression of $\Delta$Z onto Sahel precipitation (\rFigVertSections{b}) shows that these structures expand northward or shift northward at every level. The cooling of the Sahel during wet years (\rFigVertSections{a}) is, by hydrostatic balance, accompanied by a thinning of the layer below 700~hPa and an anomalous low in the mid-troposphere (\rFigVertSections{b}). The cooling of the southern part of the Saharan Low is accompanied by an increase in specific humidity (\rFigVertSections{c}) that is larger, in energy units, than the decrease in temperature, so that the low-level equivalent potential temperature, $\theta_e$, is higher over the Sahel during wet years \citep{Hurley2013}. The zonal wind (\rFigVertSections{d}) has been previously reviewed in \citet{Nicholson2013} and here we simply make the point that the northward shift and weakening in the AEJ during wet years is consistent with the mid-level anomalous cyclonic circulation over the Sahel shown in \rFigGhtMaps{d} and \rFigVertSections{d}. 

The poleward shift in the Saharan SMC can be seen as an anomalous, meridionally asymmetric quadrupole pattern in the anomalous divergence below 550~hPa (\rFigVertSections{e}; note the vertical dipole in the climatological mean fields centered at 20\dgN). There is also a meridional expansion of the upper tropospheric divergence associated with the monsoonal ITCZ. This upper tropospheric feature is also visible in the vertical velocity (\rFigVertSections{f}), with more deep ascent over the Sahel during wet years accompanying the anomalous dipole in low-level ascent that indicates a poleward shift in the Saharan SMC. Consistent with the regression of horizontal divergence, this dipole in shallow ascent is also not meridionally symmetric, with more anomalous subsidence than ascent during wet years.  There are also zonal asymmetries in the anomalies of shallow ascent, with the anomalous subsidence near 17\dgNs more pronounced in the western part of our domain and the anomalous ascent near 22\dgNs more pronounced in the eastern part (see Appendix).  

The meridional wind regression (\rFigVertSections{g}) shows an asymmetric quadrupole as well, with the southerly lobes of the quadrupole being spatially larger and of greater amplitude than the northerly lobes. However, this asymmetry in the anomalous meridional wind should not be interpreted as a weakening of the divergent circulation because we are considering a limited zonal mean, in which the non-divergent component of meridional flow does not have to equal zero. When only the divergent component of the meridional wind is considered (\rFigVertSections{h}), the meridional asymmetry in this quadrupole is substantially reduced. But to the degree that the Saharan SMC consists of both divergent and rotational components, its equatorward outflow at 700 hPa is clearly reduced during wet Sahel years (\rFigVertSections{g}).

In the next section we better quantify interannual variations in the divergent part of the circulation.  This allows us to understand how the increased mass divergence in the upper troposphere that accompanies enhanced Sahel precipitation (\rFigVertSections{e}) is  compensated by increased convergence at lower levels.  It also allows us to assess whether  asymmetries in the meridional dipoles of shallow ascent and divergence are statistically significant and thus indicate a weakening of the Saharan SMC.

\section{Changes in layer-integrated divergence}
\label{sec:results4}

\begin{figure*}
\centering
\includegraphics[width=\fullWidth]{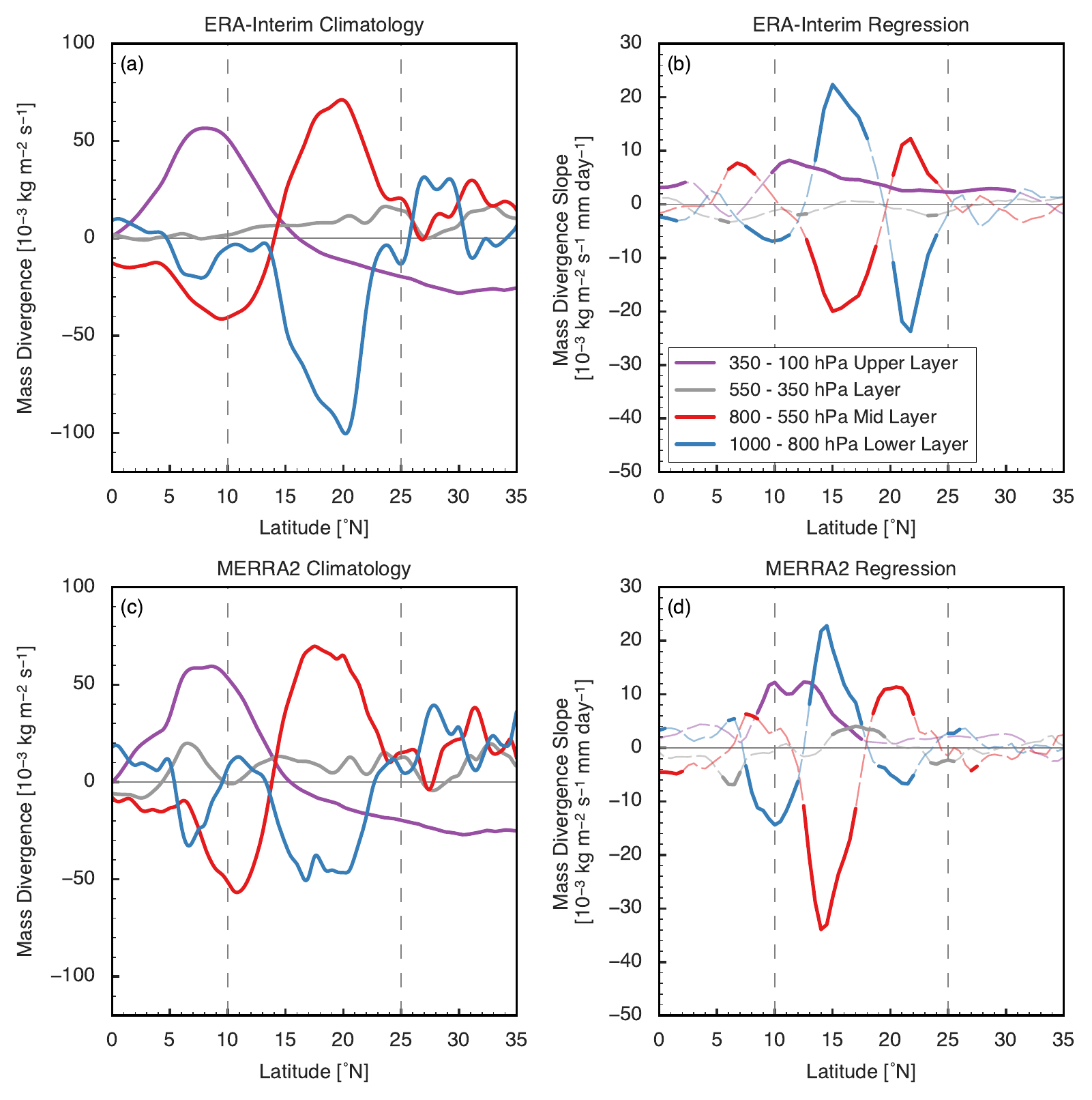}
\caption{(a) ERA-Interim climatological vertical integrals of zonally averaged divergence over the specified layers (10$^{-3}$~kg~m\mtwo~s\mone). (b) ERA-Interim regression slope of zonal and vertical integrals of anomalous divergence over the specified layers (10$^{-3}$~kg~m\mtwo~s\mone~mm\mone~day). Statistically significant slopes are shown in dark, solid colors. (c-d) As in a,b but for MERRA2. }
\label{fig:layerIntDiv}
\end{figure*}

The vertical section of divergence (\rFigVertSections{e}) allows identification of three layers that largely capture the changes in the divergent circulation: the lower troposphere (1000-800~hPa), the middle troposphere (800-550~hPa), and the upper troposphere (350-100~hPa). Taking a mass-weighted vertical integral of divergence over each layer (and over the remaining 550-350~hPa layer in which little divergence occurs) we see the climatological signatures of the divergent Hadley circulation and the Saharan SMC (\rFigLayerIntDiv{a,c}). In the lower layer, the largest convergence is due to the shallow circulation; this peaks near 20\dgN and is stronger in ERA-Interim than in MERRA2.  There is also relatively weak convergence in the ITCZ near 8\dgN.  The middle troposphere exhibits large divergence at 20\dgNs and substantial convergence at 10\dgN. Upper level divergence peaks around 8\dgN, the latitude of maximum precipitation and deep ascent;  the roughly equal magnitude and opposite signs of upper-tropospheric divergence and mid-tropospheric convergence at that latitude indicate that time-mean inflow to the deep, continental convergence zone occurs not near the surface but in the lower mid-troposphere.  There is upper level convergence poleward of about 15\dgN, consistent with the northern Sahel and Sahara being regions of time-mean subsidence. Divergence in the 550--350~hPa layer is comparatively small, and divergence above 100~hPa is smaller still (not shown). 

Regressing layer-integrated divergence onto Sahel precipitation (\rFigLayerIntDiv{b,d}) shows the now-familiar meridional dipole indicating a poleward shift of the climatological mean divergence field in the lower and mid layers, as well as a single-signed increase in upper-tropospheric divergence over the region. The asymmetry in the meridional shift is more clearly evident than in our previous depiction, with stronger changes on the equatorward lobe of the dipoles, implying that the Saharan SMC weakens as it shifts poleward. There are some differences between ERA-Interim and MERRA2, with the negative anomaly of mid-level divergence being more intense in MERRA2.  Nevertheless,  both reanalyses have asymmetric meridional dipoles in the anomalous lower- and mid-level divergence.

We wish to quantify any net change in strength of the Saharan SMC and determine if that change is statistically significant after removing the effect of the northward shift.  To do this, we horizontally average the anomalous mid-level divergence over a domain large enough to encompass the meridional dipoles in that field.  The changes in low- and mid-level divergence associated with the Saharan SMC are confined to the region between 10\dgNs and 25\dgNs (Figs.\ \ref{fig:vertSections}e and \ref{fig:layerIntDiv}b,d).  So we horizontally average the layer-integrated divergence anomalies over 10\dgN--25\dgN, 10\dgW--25\dgEs, thereby removing the antisymmetric component of the dipole and leaving a residual that corresponds to a net strengthening or weakening of divergence in each layer over the combined Sahel-Sahara region.  The 10\dgNs bound may be located slightly too far south and thus include divergence changes associated with shifts of the ITCZ, but those changes are expected to reduce the magnitude of any signal indicative of a weakening SMC (e.g.\ see the low- and mid-level anomalies in Fig.\ \ref{fig:vertSections}e). Indeed, averaging over the even larger region of 5\dgN--25\dgNs decreases the magnitude of the net divergence variations (which are presented below) but does not qualitatively change the result.

The result of this area-averaging of the layer-integrated divergence shows that during wet Sahel years, upper-level divergence increases, mid-level divergence decreases, and any changes in low-level divergence are not statistically distinct from zero (\rFigLayerDivReg{a}). Layer-averaged humidity also increases (this will be further discussed in the next section). A strengthening of the Saharan SMC would consist of a decrease in low-level divergence (enhanced convergence), and an increase in mid-level divergence; our area-averaged results have the opposite sign at mid-levels.  Furthermore, \rFigLayerDivReg{b} shows mid-level divergence is strongly anticorrelated with upper-level divergence, indicating that the enhanced upper-tropospheric divergence during wet monsoon years is balanced, in the summer-mean column mass budget, by decreased mid-tropospheric divergence.  This balance is quantitatively confirmed by the fact that the regression coefficient relating upper- and mid-level layer-integrated convergence is approximately -1.   Interannual variations in the deep, precipitating monsoon circulation thus cannot be captured by a classic first-baroclinic mode that has maximum convergence near the surface and divergence at upper levels.  \citet{Thorncroft2011} showed that the climatological mean moisture flux convergence in the Sahel has a complicated vertical structure with a weak maximum in our mid-tropospheric layer associated with flow in the Saharan SMC, so it is perhaps not surprising that variations in the flow also do not have a simple classical structure.  This issue is discussed further in the next section in the context of our idealized simulations.

\begin{figure*}
\centering
\includegraphics[width=\fullWidth]{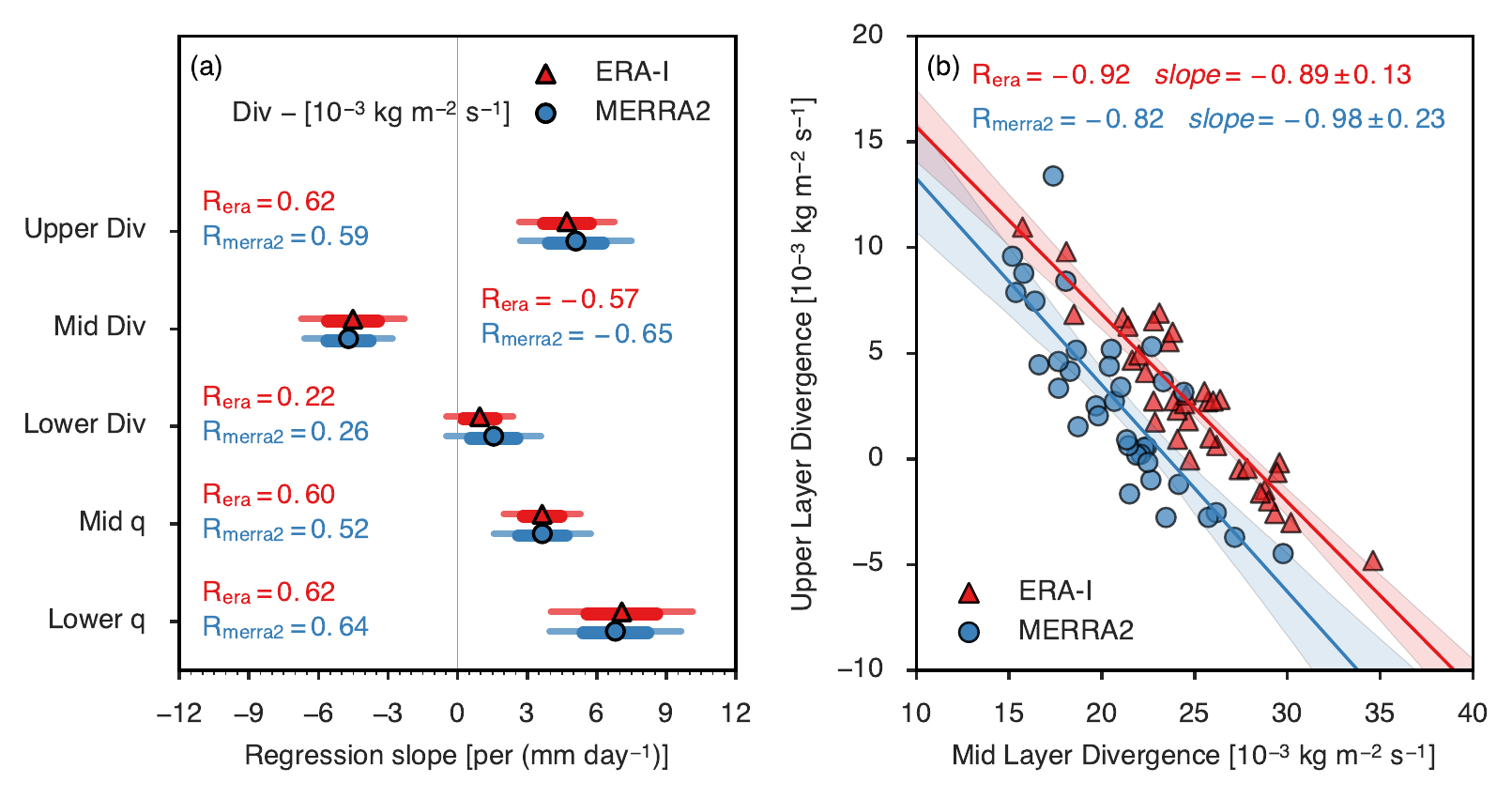}
\caption{(a) Vertically integrated divergence (10$^{-3}$~kg~m\mtwo~s\mone) or layer averaged specific humidity (10\mone~g~kg) detrended and regressed against GPCP Sahel precipitation. 68\% (thick) and 95\% (thin) confidence intervals are shown for the slope of the regression, along with the Pearson R. (b) Regression of detrended upper tropospheric divergence onto detrended mid tropospheric divergence. All quantities are horizontal averages over 10\dgW--25\dgE, 10\dgN--25\dgN.}
\label{fig:layerDivReg}
\end{figure*}

\section{Model of a weakening and shifting Saharan SMC}
\label{sec:results5}

\begin{figure}
\centering
\includegraphics[width=\halfWidth]{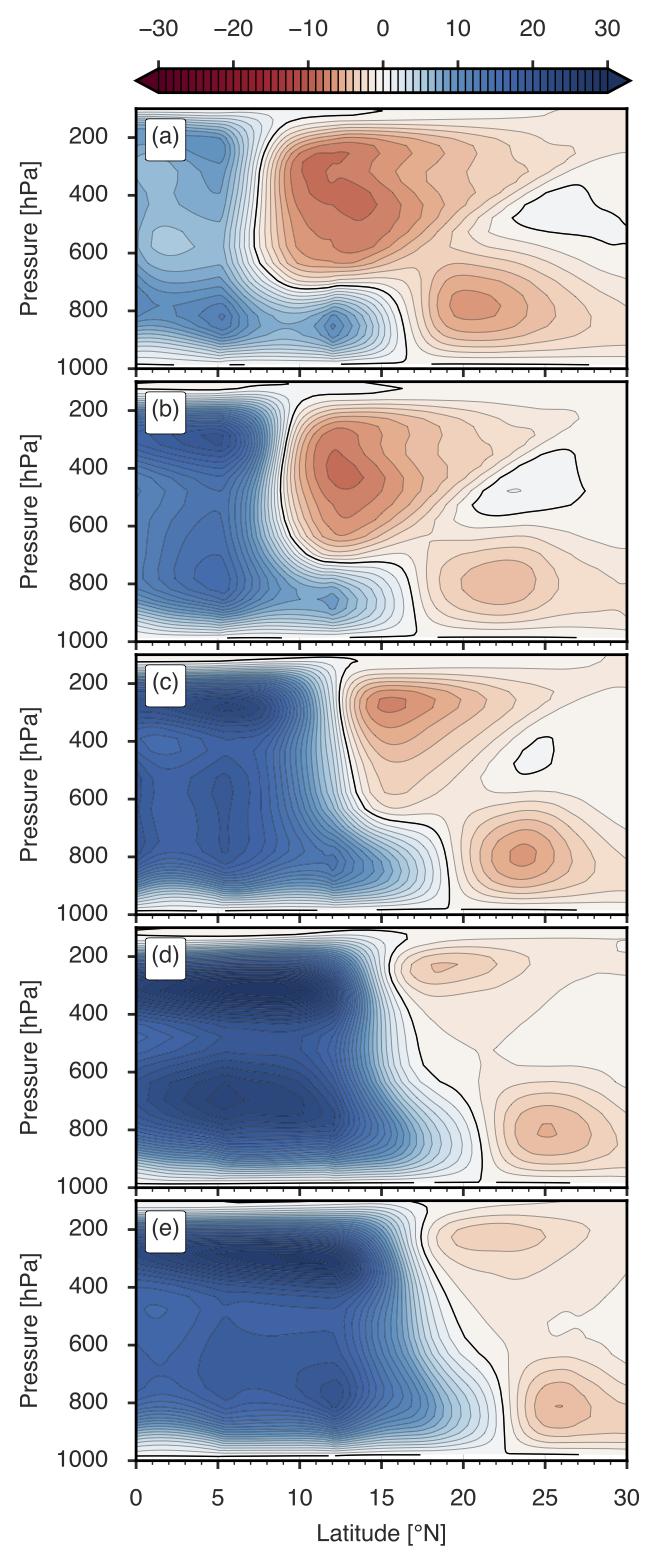}
\caption{Mass streamfunctions (10$^{9}$~kg~s\mone) ordered by increasing precipitation in the 10\dg-20\dgNs box. (a) +2K SST forcing south of Africa. (b) Albedo increase of 0.10 over 12\dg-32\dgNs. (c) Control experiment. (d) -2K SST equatorial cold tongue, as in d. (e) Albedo decrease of 0.10 over 12\dg-32\dgNs.}
\label{fig:sfnWrf}
\end{figure}

\begin{figure*}
\centering
\includegraphics[width=\fullWidth]{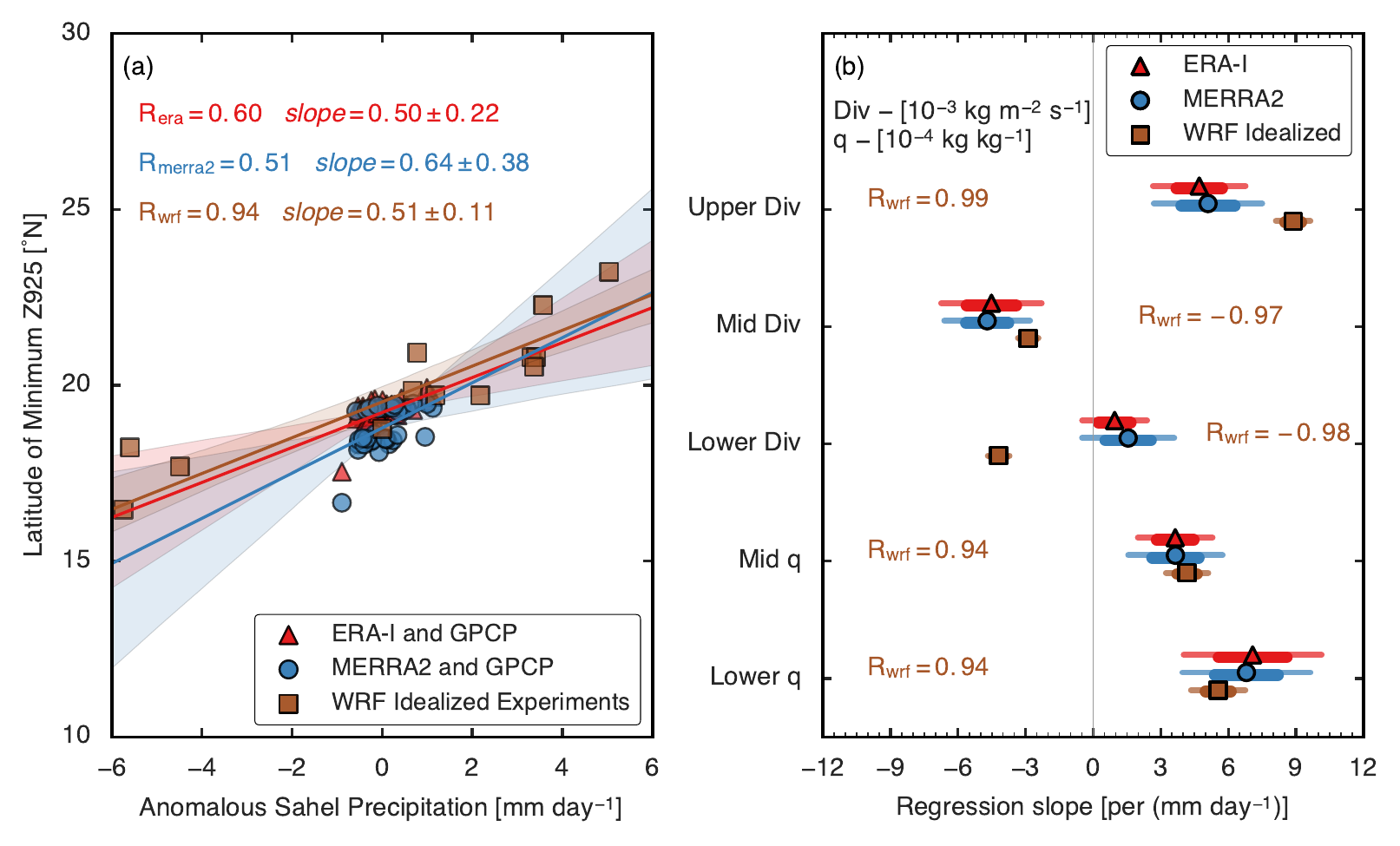}
\caption{(a) The trough location, the latitude of minimum $\Delta$Z925, was calculated and regressed against precipitation. For reanalysis data, it was regressed against detrended GPCP Sahel precipitation, and for WRF idealized simulations, it was regressed against model precipitation. (b) The layer integrated divergence (10$^{-3}$~kg~m\mtwo~s\mone) and specific humidity (10\mone~g~kg\mone) were calculated, detrended in reanalyses and regressed against GPCP, or regressed against model Sahel precipitation. 68\% (thick) and 95\% (thin) confidence intervals are shown for the regression slope. For the WRF confidence intervals, Pearson R is shown, and due to a different number of degrees of freedom (11), critical R values are 0.552, 0.683, and 0.800 at the 0.05, 0.01, and 0.001 significance levels.}
\label{fig:layerDivWrf}
\end{figure*}

Our idealized WRF model integrations, performed at 15~km horizontal resolution on a zonally periodic $\beta$-plane, are detailed in section~\ref{sec:methods} and in \citet{Shekhar2016}. A variety of surface albedo and SST forcings were applied individually about a control state to form an ensemble of model integrations. Instead of examining interannual variability within individual integrations, we look at the intra-ensemble variability of the long-term time-mean state and compare it to interannual variability within the reanalyses. Due to the zonally symmetric boundary conditions of the idealized model, the time-mean zonal wind is non-divergent and there are no large scale dynamical forcings such as those associated with ENSO or the South Asian monsoon, which could produce differences between observed interannual variability and the model intra-ensemble variability.  Nevertheless, we find quantitative similarities in the statistical association between simulated monsoon precipitation and multiple dynamical variables.

The ensemble members are strongly forced (e.g.\ Saharan albedo changes of 0.1 to 0.2), so represent a wider range of ITCZ and SMC locations than is observed in the modern historical record. After a spinup period, each ensemble member produces a different climatological state (averaged over four three-month summer seasons) in response to the applied perturbations. \rFigSfnWRF{} shows how the time-mean mass streamfunction \citep[obtained using the method of][]{Doos2011} changes between integrations having the driest Sahel and those with the wettest Sahel (the model Sahel is also defined as the region 10-20\dgN). In the ensemble member with the lowest precipitation, deep ascent peaks at 8\dgN and ascent in the SMC is well separated with a peak at 17\dgN. In this state, the summer Hadley cell is strong, the cross-equatorial winter Hadley cell is relatively weak, and the SMC  is relatively strong. As  precipitation increases over the Sahel box (e.g. \rFigSfnWRF{c}), the ITCZ moves poleward into continent, the winter Hadley cell  strengthens, the summer Hadley cell weakens, and the separation between the SMC ascent and the ITCZ decreases. As Sahel precipitation increases further (e.g. \rFigSfnWRF{e}), the winter Hadley cell continues to strengthen, the summer Hadley cell continues to weaken, and the shallow SMC ascent begins to merge with the ITCZ. 

Quantitatively comparing our idealized model with observations is complicated by the task of choosing an appropriate region over which to average precipitation and divergence.  In the reanalyses, the Sahel (10\dgN--20\dgN) always lies on the poleward edge of the ITCZ and the ascending branch of the Saharan SMC is centered in the region over which we averaged divergence (10\dgN--25\dgN).  In the idealized model, the ITCZ and SMC move over a much wider latitude band, with the ITCZ centered south of the averaging region in some integrations and squarely within it in others (\rFigSfnWRF{}).  Nevertheless, ascent in the model SMC always lies between 10\dgNs and 25\dgN, so variations in the strength of mid-tropospheric divergence produced by the Saharan SMC should be well captured by averages of the layer-integrated divergence between those latitude bounds.  For this reason, we average precipitation and layer-integrated divergence over the same regions chosen for the reanalyses.

As expected, the observed interannual variability of both Sahel precipitation and SMC latitude is much smaller than variability in the model ensemble (\rFigLayerDivWRF{a}). There is quantitative agreement between the regression coefficients based on observed and simulated variables:  the 95\% confidence interval for the slope of model Sahel precipitation regressed on model SMC latitude overlaps with that of both ERA-Interim and MERRA2.  The idealized model also exhibits associations between Sahel precipitation and upper-level divergence, mid-level convergence, and area-averaged, layer-integrated humidity that are quantitatively similar to those seen in observed interannual variability (\rFigLayerDivWRF{b}).  This agreement is remarkable given that the model was not tuned to observed interannual variability: these simulations were performed for a different study \citep{Shekhar2016} that was completed before this analysis was undertaken. However, the idealized model disagrees with observations in that it  simulates enhanced low-level convergence when there is enhanced Sahel precipitation (recall that the reanalyses indicate a weak reduction of low-level convergence during anomalously rainy years).  Enhanced precipitating ascent in the idealized model thus seems to be better described by a classic first-baroclinic mode vertical structure than in reanalyses.  Whether this means that the model is unsuitable for representing interactions between the monsoonal ITCZ and the Saharan SMC is unclear, in large part because there has been little study of the implications of deviations from a first-baroclinic mode structure for variability in monsoons. 

Despite some bias in its simulation of the vertical structure of the ITCZ, this idealized model clearly simulates a weakening and poleward shift of the Saharan SMC circulation in states with enhanced Sahel precipitation.  Furthermore, the model results suggest that the circulation over West Africa exists on a continuum.  At one end of the continuum, dry states have a coastal ITCZ close to the equator, large  separation between the ITCZ and ascent in the SMC, and a strong Saharan SMC circulation with abundant mid-tropospheric divergence.  At the other end of the continuum, the ITCZ is positioned much further poleward in a continental location, the winter Hadley cell is stronger while the summer Hadley cell is weaker, and the overturning mass flux and mid-tropospheric divergence in the Saharan SMC are weaker.  In that state, which is an idealized analogue of observed wet-Sahel years, the ascending branches of the SMC and ITCZ have begun to merge to produce a vertical structure closer to that of first-baroclinic mode ascent common in deep convective regions.

\section{Discussion}
\label{sec:discussion}

Previous work found intriguing associations between Sahel precipitation and the strength of both the Saharan Low that stretches across Africa and the SHL in the western Sahara. At interannual and longer times scales, an enhancement of either the Saharan Low or the SHL is associated with increased rainfall over the Sahel \citep[e.g.][section~\protect{\ref{sec:results2}}]{Haarsma2005, Biasutti2009,Lavaysse2015}. However, previous studies did not examine the detailed changes in the three-dimensional lower-tropospheric circulation that stretches across northern Africa --- the Saharan SMC. \citet{Martin2014a} argued that a strong springtime shallow meridional circulation across West Africa accompanied enhanced summer rainfall over the Sahel during decades when the North Atlantic was anomalously warm.  They claimed that ``the intensified shallow meridional overturning circulation increases moisture flux into the Sahel from the south during spring'', describing a mechanism in which an intensified mass flux in an SMC causes enhanced Sahel rainfall.

Here we find the changes in lower-tropospheric winds and geopotential during wet Sahel years are best described as a poleward shift and weakening of the Saharan SMC. At low levels, the decrease in geopotential was located north of the climatological mean geopotential minimum, suggesting a northward expansion or shift, rather than an intensification, of the low-level trough during wet monsoon years. When the linear relationship between the 925~hPa trough latitude and Sahel precipitation was statistically removed, effectively subtracting the poleward shift of the Saharan Low from the geopotential field, no statistically significant relationship remained between Sahel precipitation and Saharan 925~hPa geopotential. 

Our analyses of ascent and horizontal divergence showed that the divergent component of the Saharan SMC weakened and shifted poleward during wet Sahel years.  Upper tropospheric divergence over the Sahel increased during wet years, as expected for a deep, precipitating monsoon circulation.  Shallow ascent in the Saharan SMC shifted poleward and weakened during wet years, as evidenced by the meridionally asymmetric dipole of anomalous vertical velocity in the lower troposphere over the Sahara (\rFigVertSections{f}).  Asymmetric meridional dipoles were also seen in the divergence integrated over the lower and middle troposphere, confirming this weakening and poleward shifting of the shallow circulation.  The increased upper-level divergence during wet years is balanced, in the column integrated mass budget, by decreased divergence in the lower mid-troposphere, indicating some departure from classic first-baroclinic mode structures that have maximum convergence near the surface.  Nevertheless, these results suggest a trade-off  between the shallow and deep modes of vertical ascent, where unusually wet years exhibit a stronger deep circulation and weaker Saharan SMC.

An idealized model of the West African monsoon was used to produce an ensemble of integrations forced by applied SST and land surface albedo anomalies. This ensemble explores a variety of climatic states with a much greater range than that of interannual variations in reanalyses. Nevertheless, without any tuning, the intra-ensemble variability of the idealized model climatological means exhibits a similar relationship between the Saharan SMC and Sahel precipitation seen in observed interannual variability.  Increases in deep, precipitating ascent in the model were better described by a classic first-baroclinic mode than they were in reanalyses, but both the model and the reanalyses clearly showed a weakening of mid-tropospheric divergence in the Saharan SMC as monsoon precipitation increased.

All of this supports the hypothesis that the Saharan SMC inhibits, rather than strengthens, Sahel precipitation.  This is consistent with the results of \citet{Peyrille2007a}, who showed that dry and warm outflow from the Saharan high weakened Sahel precipitation in another idealized model, and that this weakening was stronger than any strengthening produced by an enhanced poleward moisture flux at low levels in the SMC.  As mentioned in the introduction, \citet{Zhang2008} also suggested that mid-level warming and drying by SMC outflow inhibits the northward progression of Sahel rainfall during early summer.  Studies of other monsoon regions have also provided evidence that a stronger SMC causes weaker monsoon rainfall:  \citet{Xie2010} established a relationship between mid-level advective drying and reduced rainfall on intraseasonal time scales in the Australian monsoon, while \citet{Parker2016} showed that the summer onset of monsoon rains is accompanied by the weakening of mid-level advective drying.  Although our observational results do not establish causation in the association between Sahel rainfall and the strength and location of the Saharan SMC, they do disprove the hypothesis that increased Sahel precipitation is caused by a strengthening of the that SMC.  In our idealized model, the ultimate cause of changes in precipitation was anomalies in SST and land surface albedo, but variations in the Saharan SMC could be part of the mechanism by which those forcings influence Sahel precipitation.

One major caveat is worth noting. \rFigOverview{c} showed that interannual variations in GPCP precipitation exhibit a ``monopole'' pattern over West Africa since 1979.  The reanalysis precipitation fields in ERA-Interim and MERRA-2 show more of a dipole pattern of interannual variability over this period, with a statistically significant decrease in precipitation over the Gulf of Guinea (5--10\dgN) during wet Sahel years (not shown).This dipole pattern of precipitation anomalies, which is consistent with a meridional shift of the ITCZ and which characterized rainfall variations earlier in the twentieth century \citep{Losada2012}, has not been seen in precipitation measurements after the 1970s. So it seems possible that the reanalyses are representing a biased spatial pattern of precipitation variability during the past few decades, assuming the precipitation observations are not themselves in error.  This uncertainty in precipitation over the coastal Gulf of Guinea region influenced our decision to use 10\dgNs as the southern boundary when calculating area averaged, layer integrated divergence, so that this uncertain region is excluded.  Even if the reanalyses have a biased representation of interannual variability in this region, it seems unlikely that this is large enough to change our broad conclusions (e.g.\ the sign of correlations between Sahel precipitation and geopotential height, divergence, and Saharan ascent would need to be wrong in two reanalysis products).  Confirmation of some of these associations in an idealized model lends further confidence to our results.  Nevertheless, it is good to bear in mind that reanalyses have bias, even while they remain a useful tool for understanding historical atmospheric variability over the last few decades.   

Important questions remain.  Does the association between a weak Saharan SMC and increased Sahel rainfall result from a time-average of a similar association on intraseasonal or synoptic time scales?  Answering this question is complicated by the fact that recent studies of subseasonal variability of the Saharan circulation have focused on selecting periods when the regional maximum of LLAT over northwest Africa (i.e.\ the darkest green shading in Fig.\ \ref{fig:ghtMaps}e) had an extreme value \citep{Lavaysse2009, Lavaysse2010, Lavaysse2010a}.  This approach tends to select periods with strong near-surface cyclonic flow west of the dateline; the concurrent precipitation anomalies are positive over the Gulf of Guinea and the central Sahel but negative over the far western Sahel.  That pattern of anomalous rainfall is distinct from the nearly zonally symmetric structure seen when the rainfall itself is used to create composites or regressions (e.g.\ Fig.\ \ref{fig:overview}c).  \citet{Vizy2014} documented an association between surges of cold air across the Sahara and reduced precipitation over the eastern Sahel on both intraseasonal and interannual time scales, but the spatial structures of the precipitation and dynamical anomalies associated with those cold surges were very different from those of the anomalies we studied here.  Although the mechanisms of subseasonal variability are important, there is a long tradition of using two-dimensional models of atmospheric meridional overturning circulations to represent the seasonal mean West African monsoon \citep[e.g.][]{Charney1975, Zheng1998}.  Our results disprove the idea a stronger Saharan SMC causes enhanced Sahel rainfall in the summer mean, and are instead consistent with the idea that the SMC inhibits Sahel rainfall by mixing dry air into that region.

\section*{Acknowledgments}
Both authors were supported by National Science Foundation grant AGS-1253222. This work was supported in part by the facilities and staff of the Yale University Faculty of Arts and Sciences High Performance Computing Center. Computing support was also provided by Yellowstone (ark:/85065/d7wd3xhc), supported by NCAR's Computational and Information Systems Laboratory. We would also like to thank Xavier Levine for many fruitful discussions. 

\appendix
\section{Appendix}

\begin{figure*}[htbp]
\centering
\includegraphics[width=\fullWidth]{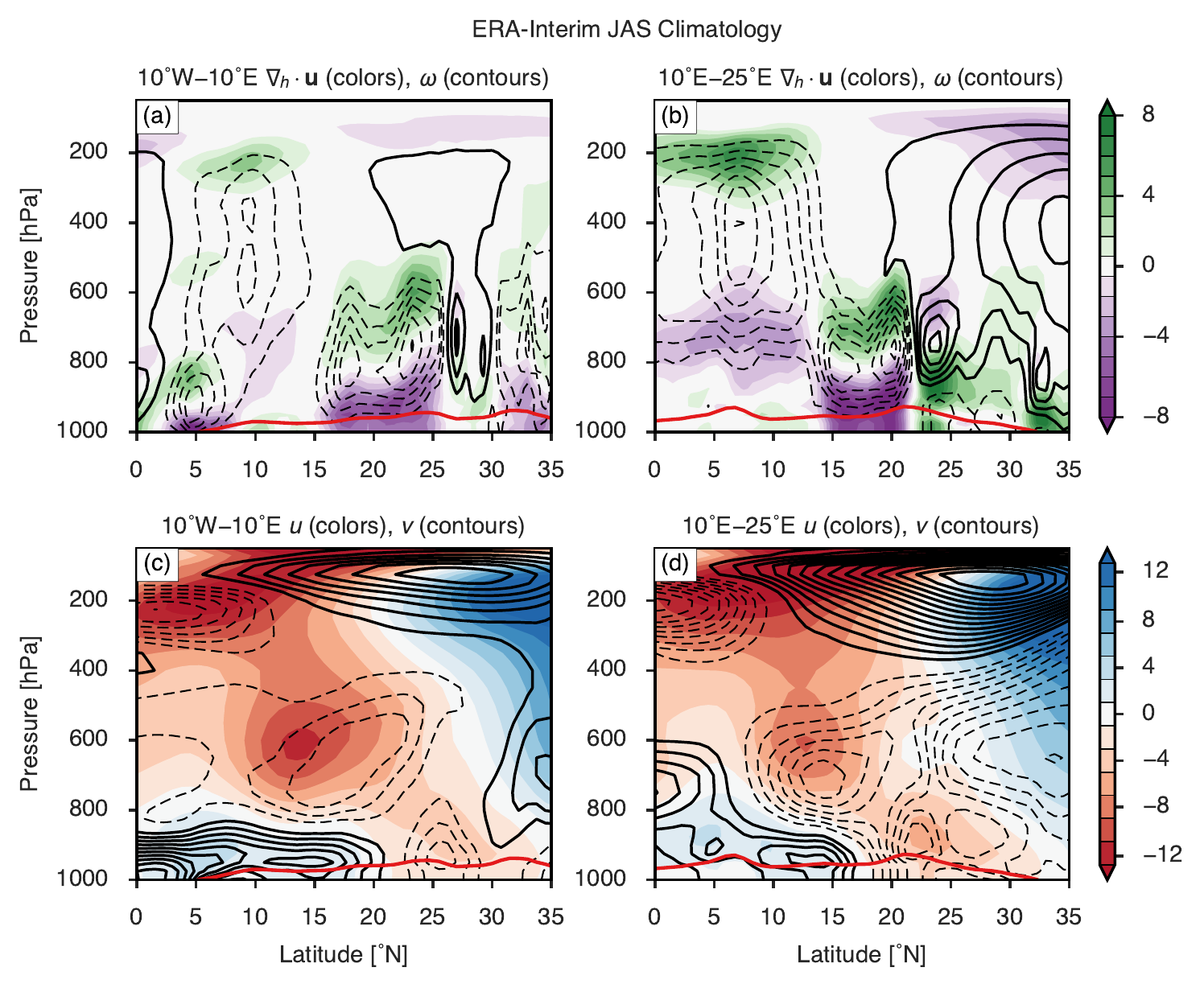}
\caption{ERA-Interim JAS Climatology zonally averaged separately in the western (10\dgW--10\dgE) and eastern (10\dgE--25\dgE) sections of the analysis domain. (a) Divergence (colors; 10$^{-6}$~s\mone) and pressure velocity $\omega$ (contours every 0.5~hPa~hr\mone) averaged over western longitudes. (b) Divergence and pressure velocity over eastern longitudes. (c) Zonal wind $u$ (colors) and meridional wind $v$ (contours every m~s\mone) over western longitudes. (d) Zonal and meridional wind over eastern longitudes. In all panels, zero contours are omitted, and negative contours are dashed.}
\label{fig:appendixSectionClim}
\end{figure*}

\begin{figure*}[htbp]
\centering
\includegraphics[width=\fullWidth]{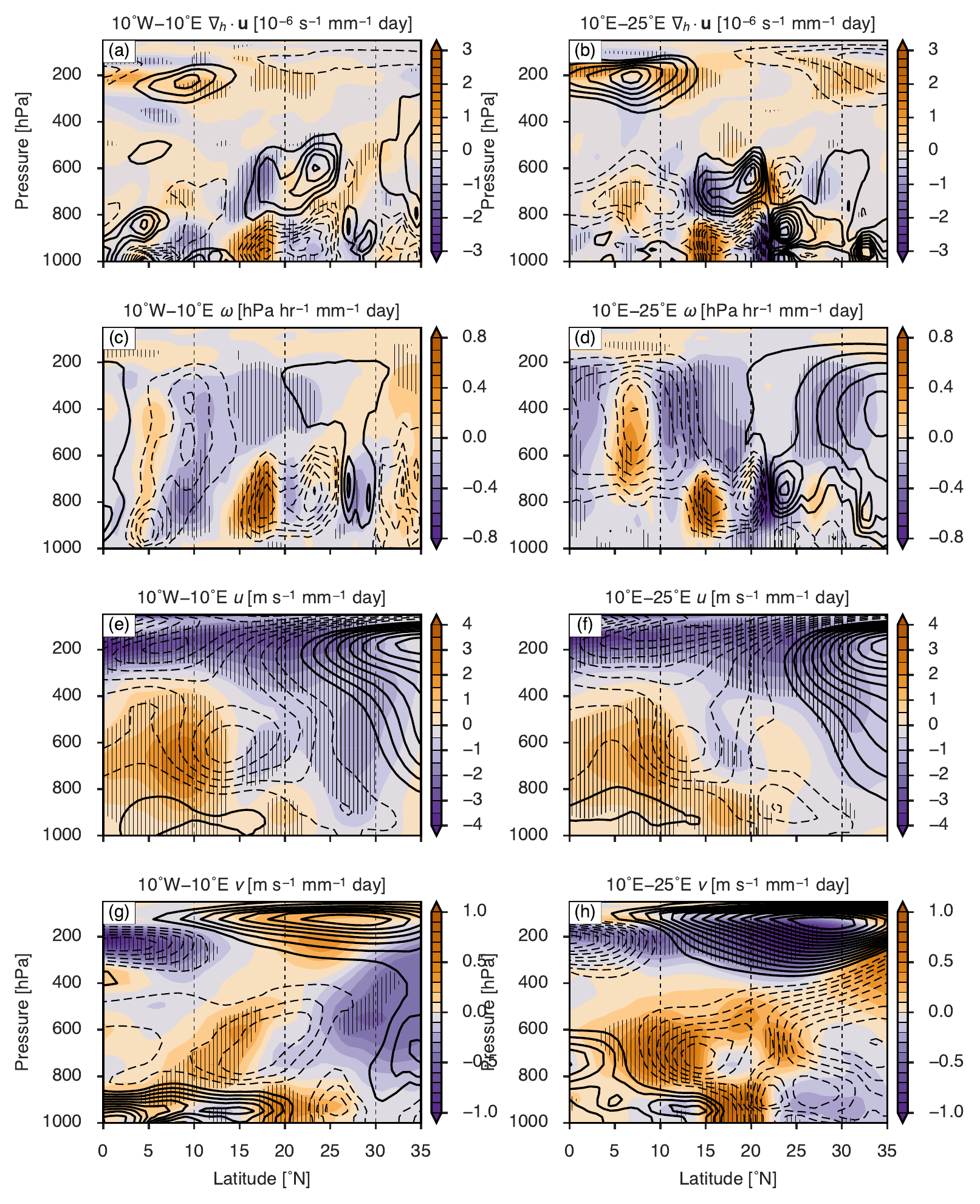}
\caption{All panels show ERA-Interim zonal averages over 10\dgW--10\dgEs (left) or 10\dgE--25\dgEs (right). Colors indicate regression slope onto Sahel precipitation, hatching indicates statistical significance, and black contours indicate climatology (negative dashed, zero omitted). Units on regression slope imply per mm~day\mone. 
(a, b) Horizontal divergence [10$^{-6}$~s\mone; contours every 1].
(c, d) $\omega$ [hPa~hr\mone; contours every 0.5].
(e, f) $u$ [m~s\mone; contours every 2].
(g, h) $v$ [m~s\mone; contours every 0.5].}
\label{fig:appendixSectionReg}
\end{figure*}

Recent work has focused on the relationship between the Saharan Heat Low in the western Sahara and its relationship to precipitation over the Sahel \citet[e.g.][]{Lavaysse2015}. In section~\ref{sec:results2}, we show there is also a region of enhanced LLAT in the eastern Sahara, extending along the 20\dgNs parallel (Fig.~\ref{fig:ghtMaps}e). Although, this feature in the eastern Sahara is not commonly called the SHL -- that term seems to be reserved for the heat low in the western Sahara --- a heat low circulation is nevertheless present in the eastern Sahara (Fig.~\ref{fig:appendixSectionClim}a, b). This circulation includes an SMC marked by near surface horizontal convergence, shallow ascent, and mid-level horizontal divergence. In the western Sahara, the shallow ascent extends from approximately 15--25\dgN, while in the eastern Sahara it is more meridionally confined, extending from approximately 14--22\dgN. 
The deep circulation also exhibits some zonal inhomogeneity:  as one moves east across Africa, the ITCZ shifts closer to the equator and is associated with more intense and meridionally broader ascent. The ITD, as measured by the zero in near-surface meridional wind (Fig.~\ref{fig:appendixSectionClim}c, d) also moves toward the equator as one moves east. This is consistent with the slight tilt of the low-level geopotential trough shown in Fig.~\ref{fig:ghtMaps}a. 

Fig.~\ref{fig:appendixSectionReg} shows dynamical fields zonally averaged over the western and the eastern parts of our analysis domain regressed on our standard Sahel precipitation time series; no meaningful differences are found when using precipitation indices derived separately from the eastern and western portions of the domain (not shown). Regressions of horizontal divergence (Fig.~\ref{fig:appendixSectionReg}a, b) show the SMC shifts north in both the eastern and western portions of the domain, but the shift is more pronounced and statistically significant in the eastern portions of the domain.  In contrast, the western part of the domain exhibits more of a weakening of the southern half of the SMC.  Regressions of pressure velocity (Fig.~\ref{fig:appendixSectionReg}c, d) show essentially the same picture, with the SMC shift showing a dipole in both eastern and western regions, but anomalous subsidence dominating the western region and ascent dominating the eastern region. This shows that if one were to consider the western region alone, the region with the SHL, wet years would be associated with a decrease in the upward shallow mass flux in the SHL. Thus, our conclusion that the SMC weakens during wet Sahel years becomes even stronger when one limits the analysis to the western part of our domain.  Averaging over a broader zonal region  makes the meridional dipole in anomalous shallow ascent (Fig.~\ref{fig:vertSections}f) show greater meridional antisymmetry.

There is a great deal of similarity between the zonal wind anomalies in the eastern and western regions (Fig.~\ref{fig:appendixSectionReg}e, f), although the barotropic subtropical jet seems to change more in the western region.  The southward wind at 700 hPa (the top of the SMC) is stronger in the eastern domain because that region lies on the eastern edge of the 700 hPa anticyclone (Fig.~\ref{fig:appendixSectionReg}g, h); the southward 700 hPa wind over the Sahel weakens more  in the eastern region during wet years, providing yet another indicator of the weakening of the entire shallow circulation over the Sahara. 

\bibliographystyle{ametsoc2014}
\bibliography{library,custom}

\end{document}